\documentclass{article}
\usepackage{amsmath}
\usepackage{caption}
\usepackage{indentfirst}
\usepackage{booktabs}
\usepackage{multirow}
\usepackage{float}
\newtheorem{rmk}{Remark}

\usepackage{titlesec}
\titlespacing*{\section}{0pt}{\dimexpr\baselineskip-20pt}{\dimexpr\baselineskip-20pt}
\usepackage{PRIMEarxiv}
\usepackage[utf8]{inputenc} % allow utf-8 input
\usepackage[T1]{fontenc}    % use 8-bit T1 fonts
\usepackage{hyperref}       % hyperlinks
\usepackage{url}            % simple URL typesetting
\usepackage{booktabs}       % professional-quality tables
\usepackage{amsfonts}       % blackboard math symbols
\usepackage{nicefrac}       % compact symbols for 1/2, etc.
\usepackage{microtype}      % micro typography
\usepackage{lipsum}
\usepackage{fancyhdr}       % header
\usepackage{graphicx}       % graphics
\graphicspath{{media/}}     % organize your images and other figures under the media/ folder

\title{\textbf{CrystalGPT}: Enhancing System-to-System Transferability in Crystallization Prediction and Control Using Time-Series-Transformers
}

\author{
  Niranjan Sitapure \\
  Dept. of Chemical Engineering \\
  Texas A\&M University \\
  College Station, TX 77801\\
  \texttt{niranjan\_sitapure@tamu.edu} \\
   \And
  Joseph Sang-Il Kwon$^{*}$ \\
  Dept. of Chemical Engineering \\
  Texas A\&M University \\
  College Station, TX 77801\\
  \texttt{kwonx075@tamu.edu} \\
}
\begin{document}

\captionsetup[figure]{font=small,skip=-15pt}
\captionsetup[table]{font=small,skip=5pt}

\maketitle

\begin{abstract}
For prediction and real-time control tasks, machine-learning (ML)-based digital twins are frequently employed. However, while these models are typically accurate, they are custom-designed for individual systems, making system-to-system (S2S) transferability difficult. This occurs even when substantial similarities exist in the process dynamics across different chemical systems. To address this challenge, we developed a novel time-series-transformer (TST) framework that exploits the powerful transfer learning capabilities inherent in transformer algorithms. This was demonstrated using readily available process data obtained from different crystallizers operating under various operational scenarios. Using this extensive dataset, we trained a TST model (CrystalGPT) to exhibit remarkable S2S transferability not only across all pre-established systems, but also to an unencountered system. CrystalGPT achieved a cumulative error across all systems, which is eight times superior to that of existing ML models. Additionally, we coupled CrystalGPT with a predictive controller to reduce the variance in setpoint tracking to just 1\%.
\end{abstract}
% keywords can be removed

\section{Introduction}
The chemical industry is undergoing a transformation with the widespread emergence of data-driven digital twins of complex chemical processes. These efforts can be categorized into three directions;  (a) Subspace identification techniques like sparse identification of system dynamics (SINDy) and operable adaptive sparse identification of systems (OASIS) that use a library of basis functions to find a sparse set of equations describing dynamic chemical processes \cite{bhadriraju2019machine,bhadriraju2021oasis}; (b) Machine learning (ML) techniques that utilize deep, recurrent or convolution neural networks (DNN, RNN, and CNN) to mimic process dynamics of crystallizers, battery systems, pulping process, catalysis, thin-film deposition, and others  \cite{sitapure2021multiscale,sitapure2021cfd,zheng2022machine, lima2022development,sitapure2022neural,hwang2022model}; and (c) Hybrid models that combine system-agnostic first principles with system-specific data-driven parameters have been demonstrated for fermentation, fracking, and other chemical processes  \cite{bangi2020deep, bangi2022physics, shah2022deep}. 

Although the aforementioned techniques show high predictive accuracy, they are fine-tuned for specific systems, resulting in poor system-to-system (S2S) transferability across different or even related systems. This limitation leads to two key drawbacks. Firstly, creating a digital twin for model prediction and control necessitates the development of a system-specific model. This requirement becomes resource-intensive for large manufacturers in chemical, food, and pharmaceutical industries that manage numerous similar chemical systems (e.g., sugar crystallizers, fermenters, pulping digesters, etc.) throughout their value chain. Additionally, the resulting collection of custom models may have different calibrations and assumptions, leading to potential compatibility issues and an increased computational burden for data management. Secondly, although customized models can be developed and maintained, the individualized training approach overlooks the substantial similarities shared by related chemical systems. For example, two sugar crystallizers may exhibit slightly different growth and nucleation kinetics, yet their interactions with manipulated variables like jacket temperature ($T_j$) and solute concentration ($C_s$) generally follow a rate equation with power law kinetics. Furthermore, the population balance model (PBM), and mass and energy balance equations (MEBEs) that describe the dynamics of these two systems are structurally similar. So while system-specific parameters may differ, these systems share significant structural commonalities defined by similar underlying equations describing state evolution. Unfortunately, these shared characteristics are not exploited during the training of individual models. Instead, each model learns a unique unified mapping function to correlate specific system input states (e.g., $T_j$, $C_s$, and others) with outputs. Given these limitations, there is a clear need for an alternative ML framework. Such a framework would leverage shared characteristics across a family of $N$ different crystal systems, thus exhibiting enhanced S2S transferability. 

To address this challenge, transfer learning (TL)-enhanced model development can be a promising avenue. Specifically, TL functions by leveraging knowledge gained from one system, and applying it to a different but analogous system. This process is based on the understanding that information about the interdependencies of process states learned from one system can be beneficial when applied to a related system. As highlighted earlier, structural commonalities, specifically in the algebraic relationship between process states, existing between Systems $A$ and $B$ can be exploited by a TL-enhanced model, simplifying its adaptation from System $A$ to $B$. Furthermore, when a specific task or system, such as online process control of a unique system, requires enhanced accuracy, the previously trained model can be fine-tuned on a smaller, system-specific dataset. In more straightforward terms, when structural similarities exist between various systems, or commonalities are present in the underlying correlations between system states, a TL-enhanced model can utilize these traits to enhance its predictive performance and reduce computational time. The primary concept here is to capitalize on the knowledge and representations acquired during training of the previous model, and further refine them to better conform to the target system. This refinement allows the model to require fewer data points about the novel and unseen system, while also reducing computational times. This is achieved in comparison to the process of training a custom-made ML model.  

Despite the current lack of a robust modeling framework with impressive TL capabilities in chemical systems, inspiration can be drawn from the disruptive emergence of large-language models (LLMs). Specifically, LLMs such as BERT, Megatron-LM, and GPT2/3/4 have led to groundbreaking applications like ChatGPT, Codex, ChatSonic, Bard, and others  \cite{vaswani2017attention,devlin2018bert,shoeybi2019megatron,radford2019language,brown2020language,liu2021swin,rombach2022high} that leverage remarkable TL abilities across different language-related tasks. The revolution in this domain has primarily been driven by the advent of transformer networks that utilize the multiheaded attention mechanism, positional encoding (PE), and parallelization-friendly architecture. These attributes expedite the development of exceptionally large models. To demonstrate, several noteworthy examples exist in terms of the use of transformer models across chemical systems. For example, Schweidtmann and colleagues converted process flow sheets into compatible tokens for a GPT2 model using the SFILES representation \cite{vogel2023learning}. this enabled the auto-completion of chemical process flow sheets. More recently, Kang and colleagues developed the MOFTransformer, which incorporates atom-based graphs and energy-grid embedding to capture both local and global features of metal-organic-frameworks (MOFs) structures \cite{kang2023multi}. Furthermore, Venkatasubramanian and Mann utilized the SMILES representation of different chemical precursors to predict potential reaction products \cite{mann2021predicting}. Thus, the potential for applying these models to new fields is substantial and promising.

Each of the above models learns from a large pool of datapoints, which are spread across $N$ different systems. These systems share certain structural similarities, such as bond patterns in MOFs or network connections between separators and reactors in flowsheets, among others. These structural similarities are captured by multiple attention heads (MAHs) connected in parallel in each encoder/decoder block. These blocks are then concatenated to produce the input/output relationship \cite{Clauwaert2021Explainability}. In simpler terms, these parallel MAHs capture different subspace dynamics of the system. When concatenated, they provide a combined representation of the overall dynamics \cite{yun2019transformers}. This approach facilitates the adaptive selection of MAH weights to yield the desired output for a specific local system, thereby demonstrating a high degree of S2S transferability. However, it is crucial to note that these models utilize static input features (e.g., protein sequence, carbon bonding, electronic distribution, etc.) to predict certain desired outputs (e.g., physical properties, selectivity, functional groups, etc.). These models do not consider the temporal evolution of system states, a factor that is essential for numerous model prediction and control tasks that are ubiquitous in chemical systems. Therefore, it becomes apparent that there is a need for the development of a next-generation digital twin. This twin should not only display high S2S transferability but should also demonstrate exceptional accuracy for time-series predictions of complex chemical systems.

\begin{figure}[!ht]
	\begin{center}
		\centerline{\includegraphics[width=1\columnwidth]{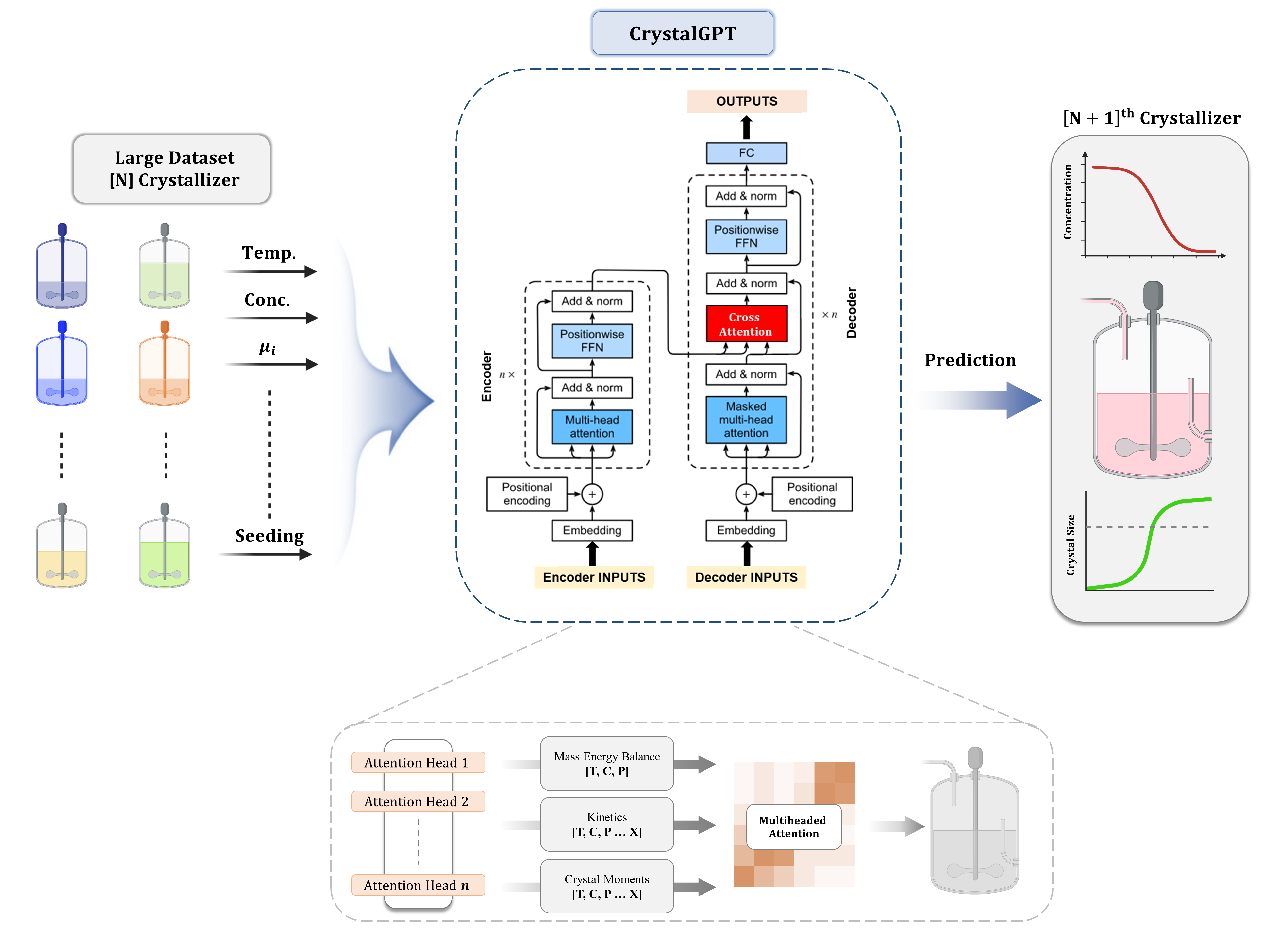}}
	\end{center}\vspace{10pt}
	\caption{Schematic illustration of the proposed CrystalGPT model.}\vspace{10pt}
	\label{transfer_learning_schematic}
\end{figure}

To address these challenges, this work presents a novel time-series-transformer (TST) framework. This framework includes variable encoder and decoder blocks, multiple attention heads (MAHs), and sinusoidal PE.  Firstly, the TST takes a $k$-dimensional state tensor, which contains information from the current and past $W$ time-steps, as an input. It then predicts a $v$-dimensional output tensor for the next $H$ time-steps, as shown in Figure~\ref{transfer_learning_schematic}. Secondly, we consider industrial batch crystallization of various sugars as a representative case study. More precisely, easily measurable process data (e.g., $T$, $C_s$, and input conditions) were collected from 20 different crystallizers, each with 5000 different operating conditions. This large dataset was used to develop a TST model (i.e., CrystalGPT) to predict the evolution of relevant system states. Thirdly, we analyzed self-attention and cross-attention scores from the encoders and decoders, respectively, to gain insights into the functioning of the TST. Fourthly, to demonstrate CrystalGPT's S2S transferability,  we tested it for model prediction and online control tasks. Specifically, CrystalGPT showcased a combined normalized-mean-squared-error (NMSE) of approximately $10^{-3}$ across all 20 crystal systems. This is eight times lower than the error of the current state-of-the-art (SOTA) long-short-term-memory (LSTM) model trained on the same dataset. Remarkably, this accuracy level was maintained when CrystalGPT was used to predict the model of a new and unseen $21^{st}$ crystal system. Additionally, we integrated CrystalGPT with a model predictive controller (MPC) for setpoint tracking of the mean crystal size ($\bar{L}$) for the $21^{st}$ crystal system. It demonstrated a setpoint deviation of $1\%$ under different operating conditions. It is important to note that the new and unseen $21^{st}$ crystal system, while not identical, is not drastically different from the previous 20 systems. Furthermore, all of these 21 crystal systems share structural similarities in their $G$ and $B$ equations, PBM, and MEBEs, allowing CrystalGPT to learn the underlying structural representations between different states. These results underline the impressive TL abilities of TSTs and highlight their potential to facilitate a high degree of S2S transferability. Moreover, they demonstrate exceptional accuracy for time-series modeling, which is an order of magnitude better than the current SOTA ML model. 

\section{Construction of Time-series Transformers (TSTs)}

\begin{figure}[!ht]
	\begin{center}
		\centerline{\includegraphics[width=0.65\columnwidth]{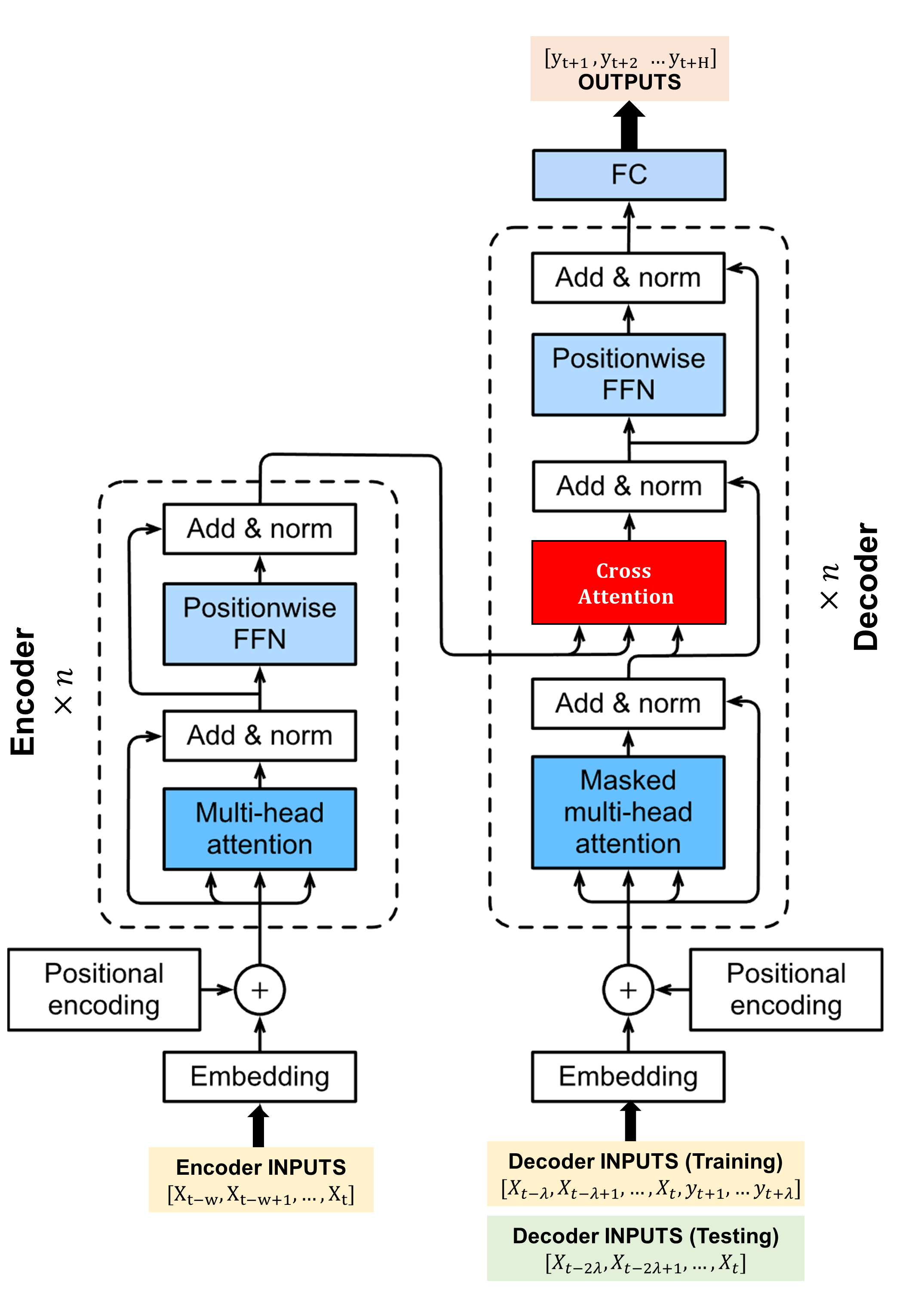}}
	\end{center}\vspace{10pt}
	\caption{Generalized architecture for an encoder-decoder TST.}\vspace{10pt}
	\label{TST_schematic}
\end{figure}

This section first provides an overview of the workings of an encoder-decoder transformer, which has served as the foundation for a majority of other transformer architectures, such as BERT, GPT, and PaLM, which use either encoder-only or decoder-only configurations. Next, a detailed description of the proposed encoder-decoder TST is presented. 

\subsection{Working of Encoder-Decoder Transformers}
Generally, a \textit{vanilla}-transformer architecture for natural language processing (NLP) tasks consists of multiple encoder/decoder blocks, each with identical sub-layers, and a globally pooled output layer  \cite{vaswani2017attention,wen2022transformers}. The input data undergoes four different transformations within an NLP-transformer. First, text inputs (i.e., sentences, paragraphs, lines of code, etc.) are preprocessed by truncation or padding of the input sequence, which is then converted into a combination of different tokens through tokenization \cite{devlin2018bert}. Second, transformer networks, unlike RNN and LSTM models, lack recurrence to explicitly consider data in a sequential manner. Consequently, they rely on the input tensor's PE to track the input's sequence. Specifically, the input tensor ($X_{RAW}$) is initially elevated into a higher-dimensional space ($d_{model}$) and processed through a sinusoidal positional encoder. This encoder embeds sine and cosine values with different phase angles to indicate the directionality and position of the words, yielding the tensor $X_{PE}$. Third, $X_{PE}$ is sent to a stack of encoder blocks. Each of these blocks employs MAHs to compute `self-attention' scores, which are processed by a feed-forward network (FFN). Each attention head calculates an attention score that captures the semantic relationship, context, and importance of each word and the entire sentence, thereby resulting in \textit{contextualized embedding}. Often, a Query-Key-Value ([\textbf{Q},\textbf{K},\textbf{V}]) approach is utilized for calculation of attention scores \cite{wen2022transformers}. Furthermore, MAHs are simultaneously trained for the same [\textbf{Q},\textbf{K},\textbf{V}], promoting automatic learning of different features from the input data. The results from all different MAHs are pooled together to yield a tensor $X_{EN}$. Fourth and most importantly, $X_{EN}$ is sent to a stack of decoder blocks, each having MAHs to compute `cross-attention' scores that are fed to a FFN. Each decoder block has its own separate input ($X_{DEC}$) and output in the form of the next predicted word in the sentence ($Y$). Specifically, $X_{DEC}$ comprises the next $H$ words in the sentence or paragraph (i.e., often referred to as the target sequence), while $Y$ represents the word with the highest predicted probability. Each decoder block computes cross-attention scores between $X_{DEC}$ (posing as \textbf{Q} values), and $X_{EN}$ (acting as \textbf{K} and \textbf{V} values). This framework allows the transformer to focus on high-value cross-attention scores, representing continuity between the source sequence and the target sequence of words, thereby generating human-like text. This method, often used in the training of transformer models, is known as the teacher-forcing method. For a more detailed description of the working of transformers, interested readers can refer to the existing literature  \cite{vaswani2017attention,devlin2018bert,sitapure2023exploring}.

\subsection{TST Architecture}

Taking inspiration from the above-mentioned encoder-decoder architecture, a novel TST framework has been tailored specifically for time-series tasks. First, it is important to note that time-series data from chemical systems consists of state information (i.e., $[X_{t-W}, X_{t-W+1} ... X_{t}]$), expressed as  \textit{float} values. Hence, a specialized tokenizer, which is commonly used to convert textual data into distinct integer-based tokens, is not necessary. Instead, the k-dimensional input tensor will be lifted to a higher-dimensional space, represented as $d_{model}$. This space corresponds to the internal dimensions of the TST. Second, the concept of positional encoding in NLP is based on the sequence of words, which does not include an explicit time component. However, in time-series modeling, a distinct time variable $t_i$ can be tracked. Consequently, the positional encoding term can be modified as follows:

\begin{equation}
	\begin{aligned}
		& Even~Position: PE_{(t_i,2j)} = sin\left(\frac{t_i}{10000^{2j/d}}\right) &\\
		& Odd ~Position: PE_{(t_i,2j+1)} = cos\left(\frac{t_i}{10000^{2j/d}}\right) & 
	\end{aligned}\label{modified_positional_encoding}\vspace{10pt}
\end{equation}
where $t_i$ is the time at location $i$ in the input sequence, $j \in \mathbb{R}^{d_{model}}$ is the feature dimension, and $10000$ is a hyperparameter. As a result of these parameters, the generated sine and cosine waves have diverse wavelengths, enabling the transformer to conveniently learn to attend to different input positions, given that the wavelength for each input value will vary  \cite{vaswani2017attention,sitapure2023exploring}. In essence, every unique input value at position $i$ and in dimension $j$ possesses a distinct identifier, that is $PE_{(i,2j})$ or $PE_{(i,2j)}$. Additionally, the wavelengths for different positions exhibit an increasing trend, providing a sense of directionality to the transformer. Broadly speaking, PE acts as a tokenizer to assign each input value at position $i$ and dimension $j$ a unique position identifier (ID). This facilitates the transformer in integrating temporal evolution into its predictions. Unlike the NLP transformer a \textit{softmax} operation at the output layer is not necessary in this scenario. This is due to the fact that the model's output for a time-series prediction task takes the form of a sequence of output variables with a prediction horizon of $H$ and dimension $v$, represented as $[y_{t+1}, y_{t+2}, ... y_{t+H}]$. The derived TST architecture is described schematically in Figure~\ref{TST_schematic}. In this representation, multiple encoder-decoder blocks are present, each encompassing \textit{{n}} attention-heads. These attention-heads accept a $k$-dimensional state tensor with an observation window of $W$ and generate a $k$-dimensional output tensor with a prediction horizon of $H$.

Furthermore, despite the similarities between the internal computation of TSTs and NLP transformers, there are several important distinctions. Specifically, in an NLP-transformer, \textbf{Q} indicates a word query (e.g., the first word of the sentence: `I like time series transformers'), whereas \textbf{K} represents all other words in the sentence in addition to the query word (i.e., [like, time, series, transformers, I]). Then, the attention is computed via a dot-product and \textit{softmax} operation, which is weighted with the original sentence (\textbf{V}) to obtain a final embedding of the sentence. On the other hand, within TSTs, \textbf{Q} symbolizes the system's state at position $i$ (i.e., $X_{t-i} \in [X_{t-W}, X_{t-W+1} ... X_{t}]$) with an observation horizon $W$ (i.e., similar to the sentence), while \textbf{K} and \textbf{V} represent the complete input tensor, $X_{PE}$. Consequently, the internal computations within an attention layer are presented as follows:  

\begin{equation}
	\begin{aligned}
		& A_{P,n} = \sum_{i}^{k} \lambda_{n,i} \textbf{V}&\\
		& \lambda_{n,i} = \frac{exp\left(\textbf{Q}^T\textbf{K}_i/\sqrt{D_k}\right)}{\sum_{j=1}^{k}exp\left(\textbf{Q}^T\textbf{K}_j/\sqrt{D_k}\right)} & \\ 
		& \sum_{i=1}^{k} \lambda_{n,i} =1& 
	\end{aligned}\label{QKV_model}
\end{equation}
where $A_{P,n}$ is the attention value for head $n$ in encoder block $P$, \textbf{Q}$\in \mathbb{R}^{D_k}$ are queries, \textbf{K}$\in \mathbb{R}^{D_k}$ stands for keys, and \textbf{V}$\in \mathbb{R}^{D_v}$ are values. Here, $D_k$ and $D_v$ are the dimensions of keys and values, respectively. Additionally, the attention score ($\lambda_{n,i}$) indicates the relative importance between different words in the input sequence, with the softmax calculation in Eq.~(\ref{QKV_model}) ensuring the scaled sum of attention scores equals 1 for each input (i.e., $\sum_{i=1}^{k} \lambda_{n,i}$ =1). It is important to acknowledge that during the self-attention process within the encoder block, \textbf{Q, K} and \textbf{V} all originate from the same input tensor, $X_{PE}$. In contrast, during the cross-attention computation within the decoder block, \textbf{Q} is derived from $X_{DEC}$, whereas \textbf{K} and \textbf{V} arise from the processed output of the encoder stack, denoted by $X_{EN}$. Furthermore, MAHs are trained on the same [\textbf{Q},\textbf{K},\textbf{V}] in order to facilitate automatic learning of diverse features from the input data. As a result, the concatenated output from MAHs represents a combination of these different features, and it is given as:

\begin{equation}
	MAH(\textbf{Q},\textbf{K},\textbf{V}) = \text{Concat}[head_1, head_2, ...,  head_n]
	\label{multiheadattention}
\end{equation}
Next, the attention scores derived from each encoder or decoder block are processed through a positional encoding, as shown below: 

\begin{equation}
FFN(\sigma_{i+1}) = ReLU(\sigma_i \theta_i + b_i)
\label{feedforward}
\end{equation}
where $\sigma_i $ represents the intermediate state from previous layers, and $\theta_i$ and $b_i$ are trainable parameters of the neural network. To improve generalization and prevent overfitting, a \textit{dropout} layer, and a \textit{Layer-norm} block can be employed between the hidden layers. A schematic illustration of the developed TST is presented in Figure~\ref{TST_schematic}. For a more comprehensive understanding of the aforementioned computations, readers are referred to the literature  \cite{wen2022transformers, zeng2022transformers, sitapure2023exploring}.

\subsection{Protocols for TST Training and Inference}
Training protocols for transformer models are different from training traditional input/output ML models due to the presence of structural differences in input and output sequences. In the training phase, the TST takes a k-dimensional tensor $X_{ENC} = [X_{t-W}, X_{t-W+1} ... X_{t}]$ as the encoder input. The decoder, in contrast, accepts a tensor $X_{DEC}=[X_{t-\lambda}, X_{t-\lambda+1}, ..., X_t, y_{t+1}, ..., y_{t+\lambda}]$, and the TST eventually outputs $[y_{t+1}, ..., y_{t+H}]$ (refer to Figure~\ref{TST_schematic}). 

In this context, $X_{ENC}$ is denoted as the source sequence, and $X_{DEC}$ is the target sequence. The term $\lambda$ represents \textit{overlap length} between $X_{ENC}$ and $X_{DEC}$. To clarify, $X_{DEC}$ consists of state values for the last $\lambda$ time-steps drawn from $X_{ENC}$ and the first $\lambda$ time-steps drawn from $[y_{t+1}, ..., y_{t+H}]$. This overlap enables cross-attention in decoder blocks, allowing them to discern the input-output relationship between $X_{ENC}$ and $X_{DEC}$. Importantly, it ensures continuity between the source and target sequences, which is pivotal for achieving high predictive performance. During the inference phase (i.e., model testing), state predictions for the next $t+\lambda$ time-steps are unavailable. Therefore, $X_{DEC}$ needs to be adjusted to include information exclusively from $X_{ENC}$. Despite the loss of overlap information, the prediction accuracy is not impacted due to the TST model's proficiency in learning the continuity and relationship of state evolution during model training. Similar protocols are observed in LLMs, wherein a forward or backward mask is used in the teacher forcing method \cite{vaswani2017attention, devlin2018bert}. The parameter $\lambda$ is another hyperparameter for the TST model, governed by certain thumb rules. It cannot exceed the prediction horizon $H$, and a $\lambda$ of zero will lead to an under-utilization of the cross-attention mechanism by neglecting overlap information. Conversely, a large $\lambda$ might cause the TST to overly rely on overlapping information, which may detrimentally affect testing results. A small $\lambda$ might not fully exploit the cross-attention mechanism. Therefore, in this study, a $\lambda$ of $H/2$ is considered. 

\subsection{Exploring Diverse Architectures of TST Models}
Given the absence of clear guidelines for selecting an optimal model size, various TST models ranging from 1 million (M) to 100M parameters were constructed, as shown in Table~\ref{TST:TST_model_archiecture}. The adopted nomenclature - [Base, Big, Large, and Mega] -  reflects the number of model parameters ($N_p$), with each model being trained on an identical dataset. Specifically, two architectural selection approaches were considered: (a) Increasing the number of encoder/decoders to deepen the TST model, thereby enhancing its capability to capture complex interactions among system states; and (b) expanding the number of attention heads, $d_{model}$, or the internal FFN's dimensionality to broaden the TST models, thereby enhancing its ability to capture numerous interactions among system states. 

\begin{table}[ht]
	\caption{\label{TST:TST_model_archiecture} Various TST model architectures considered in this work.}
	\setlength{\tabcolsep}{10pt} % Adjust the column spacing
	\renewcommand{\arraystretch}{0.75} % Adjust the row spacing
	\centering
	\begin{tabular}{@{}ccccc@{}}
		\toprule
		\textbf{}            & \textbf{Base} & \textbf{Big} & \textbf{Large} & \textbf{Mega} \\ \toprule
		Parameters        & 1.1M          & 4.5M         & 10M            & 100M           \\
		Attention heads   & 8             & 16           & 16             & 12             \\
		Head size ($d_{model}$) & 128           & 256          & 256            & 768            \\
		Encoder blocks    & 4             & 4            & 8              & 6              \\
		Decoder blocks    & 4             & 4            & 8              & 6              \\
		Neurons in FFN    & 128           & 256          & 512            & 3072           \\ \bottomrule
	\end{tabular}
\end{table}

\section{Data Generation and Visualization}

\subsection{Choosing Different Crystal Systems}

As previously discussed. the goal of this work is to develop a TST model that can effectively learn from $N$ different crystal systems. The model should provide accurate predictions across an extensive range of operating conditions and also offer baseline predictions for a new $N+1^{th}$ crystal system. Hence, we selected industrially relevant batch sugar crystallization systems as representative case studies for the training and implementation of CrystalGPT. Crystallization kinetics data for dextrose, sucrose, and lactose crystal systems  \cite{shi1990crystallization, ouiazzane2008estimation,markande2012influence} were gathered and used as a foundation for developing 20 unique synthetic crystal systems. These systems cover a broad spectrum of process scenarios and represent various polymorphs or derivative crystal systems of the traditional sugar systems. The inclusion of more than 20 unique crystal systems ensures comprehensive coverage of the parameter space, thereby allowing the TST model to discern, assimilate, and apply the structural similarities across these systems effectively. 

To this end, we formulated a generalized crystal growth rate ($G$) as follows:
\begin{equation}
	\begin{split}
		G~(m/s) = a_{\scriptscriptstyle G} \exp\left(\frac{-b_{\scriptscriptstyle G}}{RT}\right) (S-1)^{\scriptscriptstyle C_{\scriptscriptstyle G}}
	\end{split}
\end{equation}
where $R$ is the universal gas constant, $T$ is the temperature, and $S$ is the supersaturation. The kinetic parameters are denoted as $a_G$, $b_G$, and $c_G$. Similarly, the nucleation rate ($B$) is given in the following manner:
\begin{equation}
	\begin{split}
		B~(\#/kgs) = a_B M_T^{b_B}(S-1)^{\scriptscriptstyle C_{\scriptscriptstyle B}}
	\end{split}
\end{equation}
where $M_T$ denotes the suspension density and, $a_B$, $b_B$, and $c_B$ represent kinetic parameters. To construct an array of $N$ different crystal systems, these kinetic parameters (i.e., [$a_B$, $b_B$, $c_B$] and [$a_G$, $b_G$, $c_G$]) were varied according to a Gaussian distribution, as shown below: 
\begin{equation}
	\begin{split}
		p_i (x) = \frac{1}{\sigma_i\sqrt{2\pi}} 
		\exp\left( -\frac{1}{2}\left(\frac{x-\bar{p_i}}{\sigma_i}\right)^{\!2}\,\right)
	\end{split}
\end{equation}
where $p_i \in [a_B, b_B, c_B, a_G, b_G, c_G]$  represents a kinetic parameter with a mean value of $\bar{p_i}$ and a standard deviation of $\sigma_i$. For clarity, the mean value of all process parameters ($p_i$) was determined from experimentally verified values for different sugar systems. The standard deviation $\sigma_i$ was assumed to be 20\% for each $p_i$ to accommodate significant variation in growth and nucleation rates. 

\begin{figure}[!ht]
	\centering
	\includegraphics[width=0.75\columnwidth]{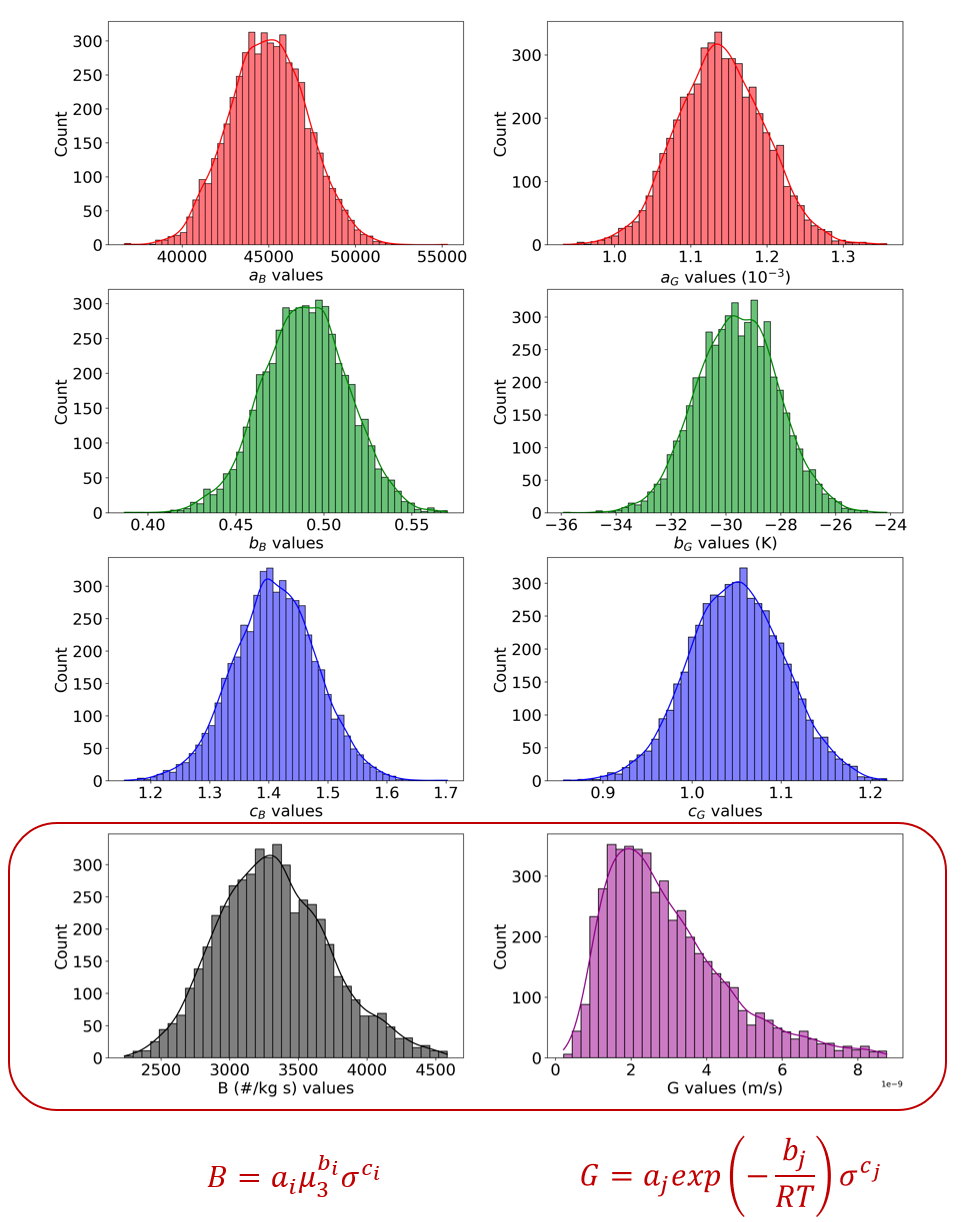}\vspace{20pt}
	\caption{The distribution of kinetic parameters ($p_i$), leading to the corresponding $G_i$ and $B_i$ values.}\vspace{10pt}
	\label{variation_G_and_B}
\end{figure}

As a result, the Gaussian distribution of each kinetic parameter, when considered together, yields a distribution of growth and nucleation rates, as described in Figure~\ref{variation_G_and_B}. Importantly, the spread of distribution in Figure~\ref{variation_G_and_B} impacts the S2S transferability, as it outlines the variable space available for the TST model to harness TL capabilities. Specifically, if we only consider 5 crystal systems from a narrow distribution instead of 20, this will hinder the TST's ability to learn transitions between different systems collectively. On the other hand, having over 50 systems, each with ample data points, enables the TST to better grasp the underlying relationship between system states, even across a broad spectrum of diverse systems. Bearing this in mind, random process parameters (i.e., $[a_B, b_B, c_B, a_G, b_G, c_G]$) for 20 distinct crystal systems, each with a unique pair of ($G_i, B_i$), were extracted from Figure~\ref{variation_G_and_B} to compile the training dataset.

\subsection{Crystallization Simulations}
\begin{figure}[!ht]
\begin{center}
	\centerline{\includegraphics[width=0.9\columnwidth]{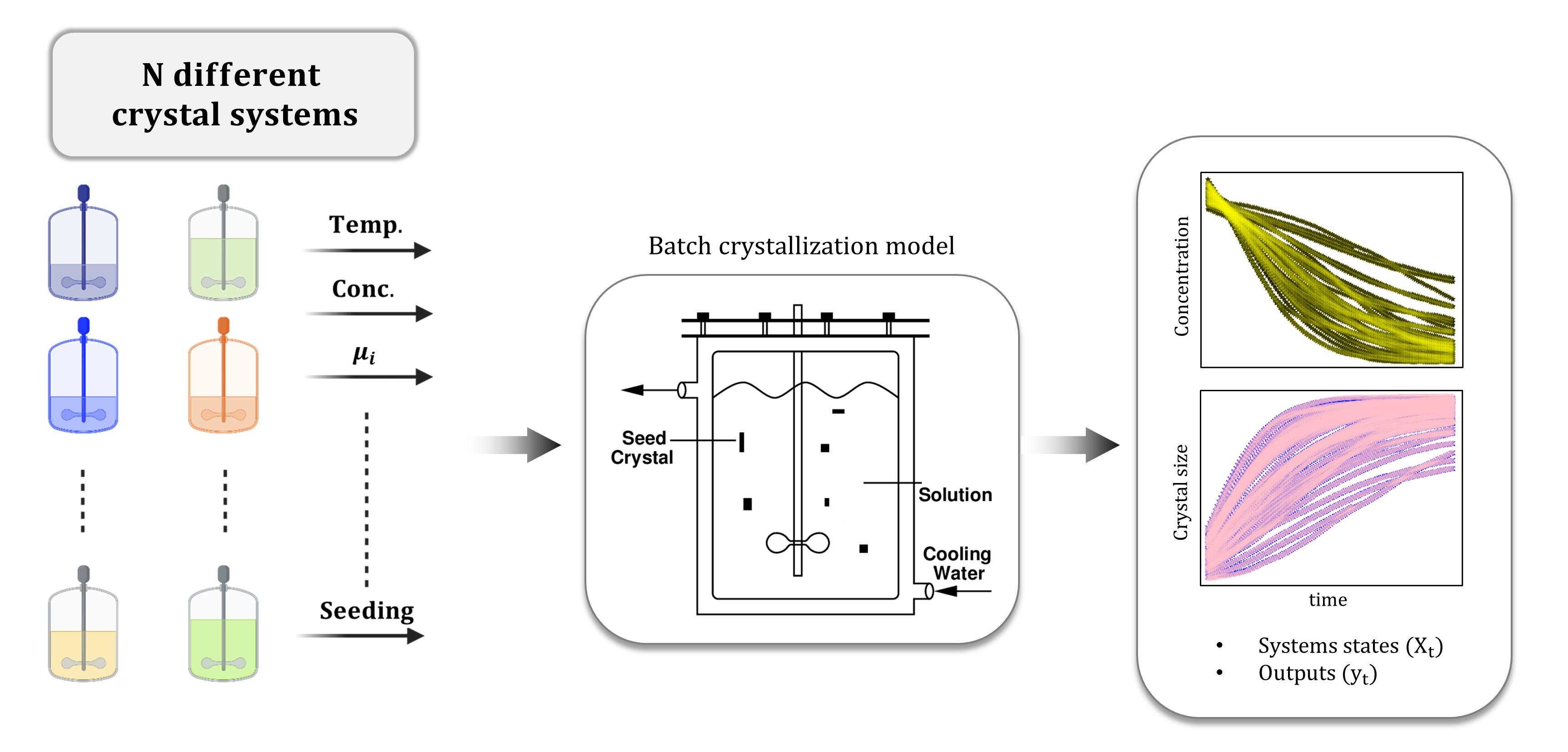}}
\end{center} 
\caption{A schematic representation of the batch crystallization process simulation.}\vspace{10pt}
\label{crystallization_schematic}
\end{figure}

For each distinct pair of ($G_i, B_i$), we consider over 5000 operating conditions, incorporating the following variations for each simulation: an arbitrarily selected (a) jacket cooling curve (i.e., $T_j (t) \in [5,40] ^{\circ}$C); (b) solute concentration (i.e., $C_o \in [0.6,0.9]$) kg/kg; and (c) seed loading (i.e., $M_{T,o} \in [0, 20]~(\%$), which is measured in\%w/w, and has a varying size for these seed crystals (i.e., $\bar{L}_o \in [10,150]~\mu m$). This extensive variation in process conditions facilitates data generation across all possible operating regimes, resulting in a comprehensive dataset that aids the TST model in effectively learning the intricate interactions between system states. Subsequently, a PBM model is integrated with MEBEs to simulate the above-mentioned 20 different crystal systems for each of the 5000+ operating conditions within a jacketed batch crystallizer. Specifically, we utilize a previously developed in-house batch crystallization simulator. Although a detailed description of the simulations can be found in our previous works  \cite{sitapure2023exploring,sitapure2023unified}, we provide a brief description here, illustrated schematically in Figure~\ref{crystallization_schematic}.

Once the desired crystallization kinetics (i.e., $G_i$ and $B_i$) have been determined, the evolution of the crystal size distribution (CSD) can be tracked using a PBM. This model employs a population density function, $n(L,t)$, as described by the following equation: 

\begin{equation}
	\begin{aligned}
		\frac{\partial n (L,t)}{\partial t} + \frac{\partial (G(T,C_s)n(L,t))}{\partial L} = B(T,C_s)
	\end{aligned}
\end{equation}
where $n (L,t)$ represents the number of crystals of size $L$ at time $t$, $B(T,C_s)$ is the total nucleation rate, and $G(T,C_s)$ represents the crystal growth rate. Subsequently, the PBM is coupled with MEBEs, as shown below  \cite{worlitschek2004model}:

\begin{equation}\label{MEBE}
	\begin{gathered}
		\frac{d C_s}{d t} = -3\rho_ck_vG\mu_2 \\
		mC_p\frac{dT}{dt} = -UA(T - T_{j}) - \Delta H\rho_{c}3k_vG\mu_2
	\end{gathered}
\end{equation}
where $C_s$ is the solute concentration, $\mu_2$ is the second moment of crystallization, $k_v$ is the shape factor, $\rho_c$ is the crystal density, and $C_p$ is the heat capacity of the crystallization slurry. Additionally, $m$ is the total mass of the slurry, $UA$ is the area-weighted heat transfer coefficient, $T_{j}$ is the jacket temperature, and $\Delta H$ is the heat of crystallization. Subsequently, the integrated set of PBM and MEBEs can be decomposed into a set of ODEs. These equations are then solved using the Range-Kutta method, implemented in Python. 

These simulations were executed in parallel using Python's \textit{concurrent futures} package. This approach generated a dataset with over 10M data points. The data was then randomly shuffled and divided into training, validation, and testing datasets using a ratio of 0.7:0.15:0.15. Furthermore, the input tensor consists of six process states (i.e., $[T_j, C_s, T,\bar{L}, M_T, time]$) for the current and preceding $W$ time-steps, while the output tensor contains four states (i.e., [$C_s, T,\bar{L}, M_T$]) for the subsequent $H$ time-steps. Further, it is important to note that monitoring instantaneous nucleation and growth kinetics for crystal systems, particularly in an industrial setting, can often be challenging. Consequently, most of the time, only directly measurable process data is available (e.g., $T_j$, $T$, $C_s$, $\bar{L}$, and $M_T$) is available. To replicate these conditions, we have designed the input tensor for the TST model to exclude any system-specific information, such as $G$ and $B$ rates. Instead, it solely comprises the aforementioned easily accessible process measurements. This tensor design is anticipated to enable the TST model to seamlessly learn to exhibit S2S transferability. In summary, by considering only observable process measurements and eliminating the need for system-specific parameters, we aim to create a model that reflects a real industrial setting. 

\begin{rmk}
It is worth noting that sugar crystallization systems (e.g., Dextrose, Maltose, Erythritol, Sorbital, and others) are prevalent in the food industry. Crystallization kinetics of similar sugar molecules (e.g., Maltose, Erythritol, Sorbital, and others) will not differ drastically. Furthermore, these sugar systems can be adequately characterized by similar structures of $G$ and $B$ equations, PBM, and MEBEs, as outlined in the previous section.  Therefore, these crystal systems can be grouped together as they exhibit similarities without drastic differences. This allows for the formulation of a TST model specific to this cluster of sugar crystallization systems. However, it is important to note that this model may not be applicable to starkly different systems such as quantum dot (QD) systems, which exhibit ultra-fast process dynamics, the presence of different ligands influencing surface growth, among other factors \cite{epps2017automated, dong2018precise, sitapure2020kinetic}. Thus, when developing such a model to enhance S2S transferability, it is essential to consider a family of crystal systems that share structural similarities and process parameters derived from a similar distribution (as depicted in Figure~\ref{variation_G_and_B}). 
\end{rmk}

%\subsection{Process Description}
%
%The overall construction of CrystalGPT is schematically described in Figure~\ref{transfer_learning_schematic}. Firstly, often a certain chemical plant in the food for the pharmaceutical sector will have $N$ number of crystallizers, each with slightly different or starkly different crystal systems (e.g., lysozyme and paracetamol in pharmaceutical companies vs. dextrose, maltose, and Erythritol in food companies). Secondly, for each $N$ crystallizer, time-series data of system states (i.e., temperature, solute concentration, crystal moments, etc.) can be extracted through the plant monitoring or control system that have access to various sensor measurements, which can result in a humongous corpus of rich process data. Unfortunately, either the current practices do not make use of this data or various bespoke ML models (i.e., disparate models for each crystal system) are generated, thereby not fully realizing the potential of such a large data repository. To overcome this issue, thirdly, a large transformer model (i.e., CrystalGPT) is formulated to process the entire corpus of process data from $N$ crystallizers and is subsequently trained for 10 to 25 epochs. Finally, the resulting TST model is trained to provide accurate test predictions for any randomly chosen crystal system with arbitrary operating conditions. 

\begin{figure}[!ht]
	\begin{center}
		\centerline{\includegraphics[width=0.8\columnwidth]{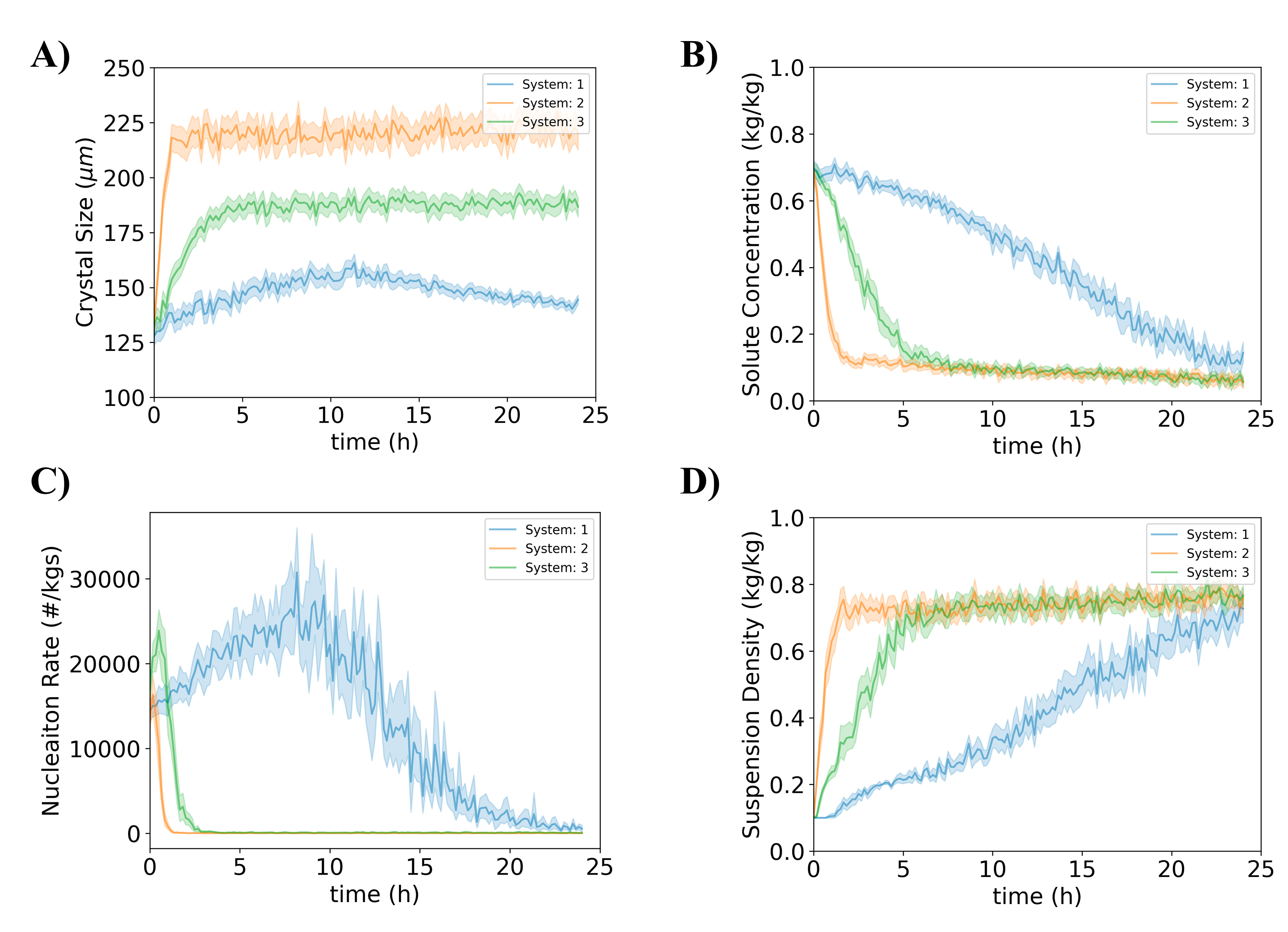}}
	\end{center} \vspace{10pt}
	\caption{Compilation of system state evolution for three representative crystal systems.}\vspace{10pt}
	\label{compilation_different_crystal_systems}
\end{figure}

\subsection{State Evolution in Representative Crystal Systems}
The visualization of data from numerous diverse systems and operating conditions can be cumbersome, potentially obscuring the differences between simulations. Thus, Figure~\ref{compilation_different_crystal_systems} selectively presents three distinct crystal systems out of the complete set of 20. These systems were specifically chosen for their representative qualities, each exhibiting differing nucleation kinetics but comparable growth kinetics. Our aim was to highlight how a single system-specific parameter can influence the development of various crystal systems, a concept our newly developed TST model is designed to comprehend.  

To this end, despite operating under similar conditions, these three systems display notably different system state evolution (i.e., mean crystal size, concentration, and suspension density) as seen in Figure~\ref{compilation_different_crystal_systems}. The striking variations underscore the role of differing nucleation rates ($B$), even under similar operating conditions. Specifically, System 1 exhibits sustained nucleation kinetics, peaking around the midpoint of operation (approximately at 10 hours), leading initially to an increase and subsequent decrease in mean crystal size due to the creation of numerous fine crystals. This system also demonstrates a steady increase in overall suspension density coupled with a consistent decrease in solute concentration. In contrast, Systems 2 and 3 show rapid initial crystal growth accompanied by relatively weak nucleation rates, resulting in stabilized mean crystal size and suspension density. The gradual subsequent growth can be attributed to a drop in solute concentration (i.e., a lower driving force). Overall, Figures ~\ref{compilation_different_crystal_systems} and \ref{variation_G_and_B} showcase the challenges an ML model would face in mirroring such a diverse dataset. It is particularly daunting to accurately predict  time-series data across 20 different systems using only observable process variables (e.g., $T_j$, $T$, $C_s$, $\bar{L}$, and time), given the substantial influence of system-specific parameters, such as the nucleation rate.

\section{Results and Discussion}

This section serves two primary purposes: (a) it presents model testing results to highlight the S2S transferability of CrystalGPT, and compares these findings with the SOTA LSTM model, and (b) it examines two case studies (i.e, model prediction, and MPC implementation) to underline the practical applicability of this model. 

\subsection{S2S Transferability}
A training, validation, and testing dataset, containing roughly 10M datapoints randomly selected from 20 different systems under varying operating conditions, was generated. The NMSE values for different CrystalGPT models are outlined in Table~\ref{CrystalGPT: table}. First, all the models show a low NMSE value of approximately $5\times10^{-4}$ to $10\times10^{-4}$ for both $W=12$ and $W=50$ cases. Furthermore, expanding the size of the TST has only a marginal impact on NMSE value up to a certain point. Beyond that, for particularly large model sizes (e.g., Mega model), the predictive performance declines. This phenomenon is attributable to an optimal ratio that should be maintained between the number of model parameters ($N_p$) and the size of the dataset ($\mathbb{D}$) to achieve a desirable predictive performance \cite{sitapure2023exploring}. 

\vspace{10pt}
\begin{table}[!ht]
	\centering
 \caption{Performance comparison of various CrystalGPT models and LSTM: Best-performing models for each case is highlighted in \textbf{bold}.}
	\renewcommand{\arraystretch}{1} % Adjust the value to reduce line spacing
	\begin{tabular}{@{}ccccccc@{}}
		\toprule
		&            & \textbf{LSTM} & \textbf{Base} & \textbf{Big} & \textbf{Large} & \textbf{Mega} \\ \bottomrule
		\# Parameters                                                                      &            & 125K          & 1.1M         & 4.5M             & 10M                & 100M              \\
		&            &               &              &                  &                    &                   \\
		\multirow{3}{*}{\begin{tabular}[c]{@{}c@{}}NMSE ($10^{-4}$)\\ $W$ = 12\end{tabular}} & Training   & 34.51          & 8.93          & 8.93              & \textbf{8.27}       & 9.38               \\
		& Validation & 34.76          & 9.31          & \textbf{9.01}     & 9.14                & 9.68               \\
		& Testing    & 35.01          & 9.33          & \textbf{9.29}     & 9.35                & 9.69               \\
		&            &               &              &                  &                    &                   \\
		\multirow{3}{*}{\begin{tabular}[c]{@{}c@{}}NMSE ($10^{-4}$)\\ $W$ = 50\end{tabular}}   & Training   & 24.02          & 5.25          & 6.13              & \textbf{4.17}       & 6.38               \\
		& Validation & 24.44          & 6.21          & 6.55              & \textbf{6.10}       & 6.58               \\
		& Testing    & 25.54          & 6.3          & 6.61              & \textbf{6.15}       & 6.62               \\ \bottomrule
	\end{tabular}\vspace{10pt}
	\label{CrystalGPT: table} 
\end{table}

The next step involves comparing our approach with the current SOTA ML models. To this end, we trained a 4-layered LSTM network, consisting of 512 LSTM cells, on the same dataset. LSTM was selected as it represents a specialized version of sequential RNN models, specifically designed to handle time-series data. More specifically, RNNs incorporate an internal \textit{for} loop to process the state value from the preceding time-step (i.e., $[X_{t-W}, X_{t-W+1} ... X_{t}]$). However, due to this explicit consideration of the source sequence, RNNs often suffer from high training times and issues with gradient vanishing or exploding. In contrast, LSTM is both computationally efficient and highly accurate. It employs a combination of input, output, forget, and update gates to selectively transmit pertinent information to the next layer. Consequently, LSTM has been widely recognized as a SOTA technique for time-series modeling within the literature. Our results clearly indicate that all CrystalGPT models show approximately 5 to 8 times lower NMSE values than LSTM. This significant performance improvement is noteworthy, especially considering that LSTM explicitly accounts for all time-steps in a source sequence, while TSTs use an indirect method of positional encoding and a simple FFN for their internal computations. Moreover, it was observed that for both LSTM and TST models, expanding the window size, $W$, provides better predictions. This outcome is due to the fact that longer source sequences contain more detailed information about the dynamics of different states, thereby enabling the ML model to capture more complex interdependencies between different states \cite{sitapure2023exploring}. 

\subsection{Testing on $N+1^{th}$ Crystal System}
The results presented thus far highlight CrystalGPT's adeptness in accurately characterizing a set of 20 distinct crystal systems across a broad spectrum of operating conditions. To further validate CrystalGPT's S2S transferability, we tested its performance on a completely new, unseen $21^{st}$ crystal system. A new crystal system, unique in its growth rate ($G_i$) and nucleation rate ($B_i$) values, was simulated under 2500 operating conditions. This $21^{st}$ crystal system was created by drawing a new set of parameters (i.e., $[a_B, b_B, c_B, a_G, b_G, c_G]$) from the distributions depicted in Figure~\ref{variation_G_and_B}. In essence, this $21^{st}$ System aligns structurally with the previous 20 systems in terms of the forms of $G$ and $B$, PBM, and MEBEs, but it carries distinct system parameters. 
We considered two scenarios for this evaluation: (a) baseline predictions from CrystalGPT without any fine-tuning (i.e., without additional training on new data from the $21^{st}$ System), and (b) fine-tuning CrystalGPT with process data from the $21^{st}$ System, and then testing its predictive performance. Further, it is important to note that only 2500 operating conditions for the new $N+1^{th}$ System were considered because of two reasons; First, since we are interested in testing CrystalGPT’s predictive performance on an unseen and new $N+1^{th}$ crystal system, we only need to a testing dataset instead of a larger dataset that can be split into training, validation, and testing datasets. Second, we are also interested in analyzing the effect of model fine-tuning for a new N+1 system, which would require the procurement of an adequate amount of training data. However, since CrystalGPT has already been trained on a comprehensive dataset of 20 different systems, having fewer datapoints for the fine-tuning tasks should suffice. Thus, accounting for the above factors, in the case of $N+1^{th}$ System, simulation data for 2500 (instead of 5000) different operating conditions were considered. 

\begin{table}[!ht]
	\centering
	\caption{A comparative analysis between the baseline predictions and fine-tuned predictions for the $21^{st}$ crystal system.}
	\label{baseline_table}
	\renewcommand{\arraystretch}{1} % Adjust the value to reduce row spacing
	\begin{tabular}{@{}cccccc@{}}
		\toprule
		\multicolumn{1}{c}{} & \textbf{NMSE ($10^{-4}$)} & \textbf{LSTM} & \textbf{Base} & \textbf{Big} & \textbf{Large} \\
		\midrule
		\multirow{2}{*}{W=12} & \textit{Baseline} & 10.72 & 1.56 & 1.46 & \textbf{1.13} \\
		& \textit{Fine-tuning} & 9.05 & 1.15 & \textbf{1.05} & 1.09 \\
		\midrule
		\multirow{2}{*}{W=50} & \textit{Baseline} & 9.74 & \textbf{1.45} & 1.52 & 1.54 \\
		& \textit{Fine-tuning} & 8.85 & 1.04 & 1.09 & \textbf{0.98} \\
		\bottomrule 
	\end{tabular} \vspace{10pt}
\end{table}

Table~\ref{baseline_table} presents the NMSE values for CrystalGPT, both for the baseline scenario and the fine-tuned case. From the results, it is evident that CrystalGPT consistently maintains high prediction accuracy in both scenarios. Mirroring its performance with the initial 20 systems, CrystalGPT outperforms the LSTM model by a factor of 8 to 10, delivering low NMSE values around $10^{-4}$. Remarkably, even in baseline scenarios without prior exposure to the new crystal system, CrystalGPT yields impressively low NMSE values ($1.5\times10^{-4}$), thus demonstrating its robust transfer learning capabilities. This proficiency can be attributed to CrystalGPT leveraging the shared structural features of the PBM and MEBEs that describe both the previous 20 systems and the new, unseen $21^{st}$ System. What is more striking is the fact that these NMSE values are compiled from over 750 operating conditions and their corresponding data-points, highlighting CrystalGPT's ability to maintain high accuracy across a broad range of operations for an entirely new and unknown system, even without any fine-tuning.  

\begin{rmk}
Furthermore, the model can be fine-tuned with a small amount of data from the $21^{st}$ System. However, the difference in NMSE values between the baseline and fine-tuned scenarios is trivial, thereby suggesting that additional fine-tuning is not required in practice. This applies to an $N+1^{th}$ crystal system, possessing system parameters (i.e., $[a_B, b_B, c_B, a_G, b_G, c_G]$) that are part of the distribution displayed in Figure~\ref{variation_G_and_B}. CrystalGPT can leverage shared structural similarities among the PBM, MEBEs, and kinetics, thus demonstrating seamless capabilities without the necessity of further fine-tuning. On the other hand, for an $N+P^{th}$ crystallization process, if the system parameters markedly diverge from those presented in Figure~\ref{variation_G_and_B}, CrystalGPT might not deliver accurate baseline predictions. In such an instance, it becomes essential to fine-tune the model using additional process data for the $N+P^{th}$ System. This shared structural commonality between the PBM, MEBEs, and kinetics of the preceding $N$ systems and a newly unseen $N+P^{th}$ System can potentially empower CrystalGPT to expedite its training process. 
\end{rmk}

\subsection{Multiheaded Attention Empowers S2S Transferability}

\begin{figure}[!ht]
	\begin{center}
		\centerline{\includegraphics[width=1\columnwidth]{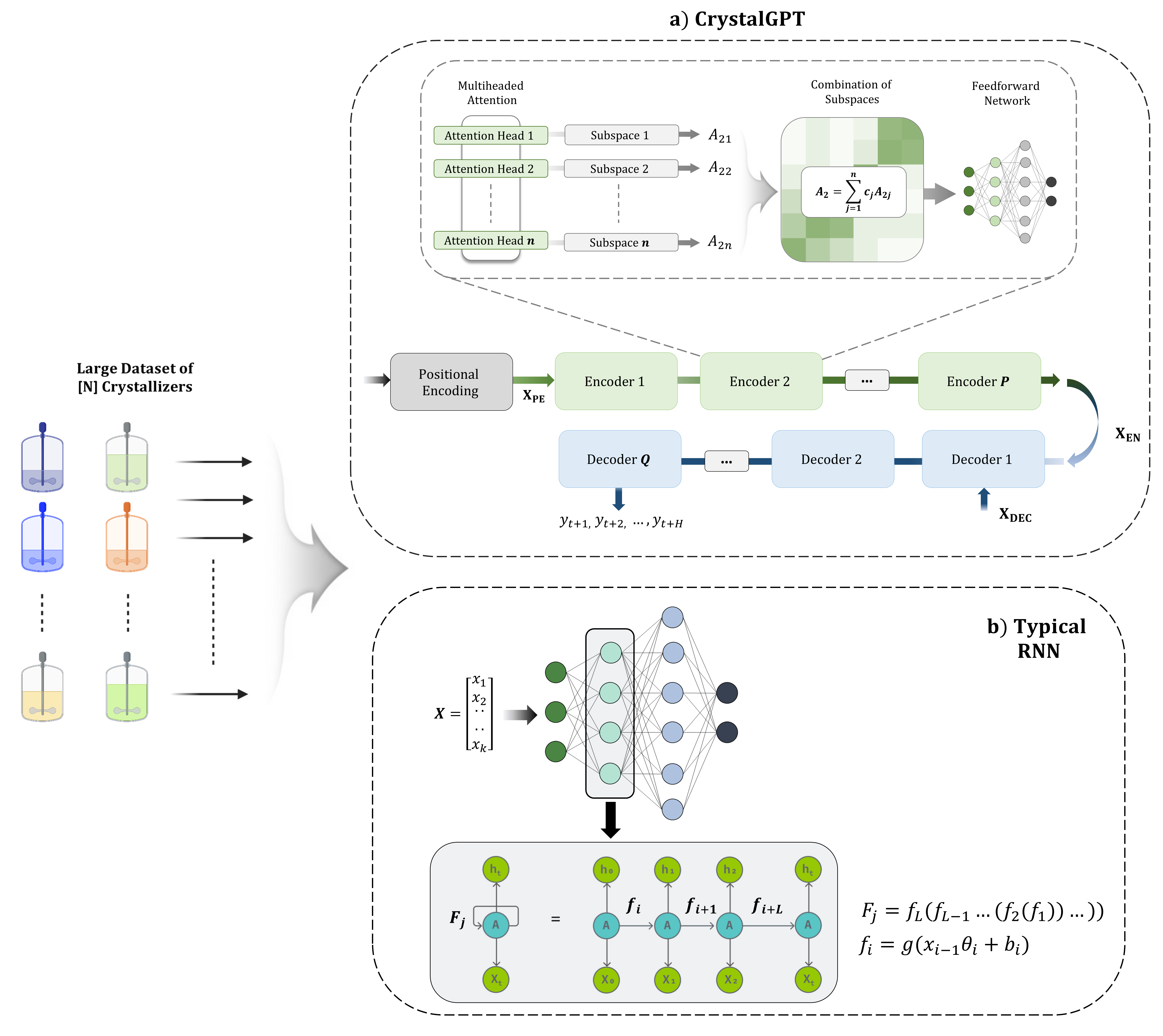}}
	\end{center} \vspace{10pt}
	\caption{Illustration of the functional differences between LSTM, a type of RNN, and CrystalGPT.} \vspace{10pt}
	\label{TST_inner_working}
\end{figure}

The aforementioned results highlight the superior performance of CrystalGPT when compared to SOTA LSTM, a type of RNN model. However, it is important to understand the distinct attributes of a TST-based model that enhance its generalization capacity over $N$ or $N+1$ different systems. Figure~\ref{TST_inner_working} presents a schematic comparison between a TST model and a typical RNN. Primarily, three fundamental aspects within TSTs contribute to their improved predictive performance.

First, each encoder and decoder block has $n$ attention heads, followed by an FFN. The same input tensor is fed to each attention head, which then computes the attention values $A_{ij}$ for attention head $i$ in encoder/decoder block $j$. Subsequently, attention values from all the heads are aggregated ($ \sum^{n}_{j=1} c_j A_{ij}$) using trainable network parameters $c_j$ and passed through an FFN with a ReLU or sigmoid activation function to approximate an input/output correlation. This procedure is repeated for each successive encoder/decoder, yielding complex representations of the input/output relationship. For any System $K \in [0, N]$, the parallelized nature of MAHs, which tends to have the same input tensors, allows the TST to dissect a unified mapping function between inputs and outputs into multiple constituents (i.e., subspaces). These can then be collectively weighted to represent the unified mapping function. However, when the TST is trained with input data for $N$ systems, each possessing $k$ features, each individual head `attends’ to different interdependencies among the various system states. This breaks down the unified mapping function into not just its constituents but also shared subspace models between different system states. Examples include the relationship structure between $T$ and $C_s$ or $C_s$ and $\bar{L}$, and others. When combined, there can collectively represent the crystallization dynamics. Furthermore, during model training, the TST model parameters can learn to adapt the weights of the attention values ($c_j$) to various local regions within the entire state space. For example, input state information from System $A$, which is part of a pool $N$ systems, is associated with the attention scores' weights within a specific local region. On the contrary, a different region would correspond to System $B$, another member of the same pool of $N$ systems. As a result, the TST model learns to establish a map between different systems, associating them with corresponding attention scores' weights.

Second, the distributed internal framework described above is crucial for maintaining high accuracy when extending the TST model to a new, $N+1^{th}$ System. When the state information of an unfamiliar system is input into the TST, despite not having encountered such information during training, the various attention heads within the model are able to identify and focus on the structural similarities between the interdependencies of the system states. Subsequently, these can be combined using a unique set of weights ($c_j$) to accurately depict the state dynamics for the $N+1^{th}$ System. For example, consider a hypothetical TST. One attention head may capture the relationship between temperature ($T$) and solute concentration ($C_s$), represented by a solubility curve. Another attention head might capture the correlation between the growth rate $G$ and $C_s$ (i.e., the growth rate equation). During training, the TST model can learn these distinct representations along with the various possible combinations of attention values from different heads. When exposed to state information from a new system, which is structurally akin to the previous $N$ systems, the TST model can compile attention scores from $n$ different heads for $P$ encoders and $Q$ decoders. This process yields a unique amalgamation of all attention values that can accurately describe the dynamics of the $N+1^{th}$ System. 

Third, the capabilities described above can be further enhanced by adjusting the width and depth of the TST \cite{Levine2020Limits}. The width of a transformer model is majorly influenced by the number of attention heights and the size of the FFN. A larger number of attention heads allows for the capture of prominent and subtle system interdependencies, which can be pooled together for an improved understanding of the input/output relationship. However, the performance improvement gained by increasing the number of attention heads will eventually plateau. This occurs when the existing $n$ number of attention heads already represents the dominant modes of system dynamics, resulting in diminishing returns for additional heads. Conversely, the depth of a TST model is determined by the number of encoder/decoder blocks. Given that every encoder/decoder block consists of $n$ attention heads in parallel, which are connected to a simple FFN layer, increasing the number of encoder/decoders allows each successive FFN to capture more nonlinearity in the input/output relationship  \cite{Xu2019Lipschitz, song2020specswap}. In conclusion, the width and depth of TSTs are critical hyperparameters that significantly influence the TST model's S2S transferability.

In contrast, an RNN is made up of multiple fully connected layers of recurrent neurons. Specifically, the output from each recurrent neuron can be represented as $F_j$, which is a convolution of $L$ different neurons. Each neuron has an activation function $f_i$ that represents $L$ time steps from the source sequence. During the training a RNN for a given $K^{th}$ System, the model training aims to find a single unified mapping nonlinear function ($\phi$) that can describe the correlation between inputs and outputs. Although this is extremely powerful for a specific System $K$, extending the model training to $N$ different systems makes it highly challenging to procure a single unified mapping function $\phi$. Furthermore, the RNN model considers all the time-steps in a sequential manner, giving equal importance to all the state inputs. This approach inefficiently handles temporal information, leading to the issue of vanishing/exploding gradients over large window sizes. Given these two features, if such an approximate function is computed by a certain RNN model, extending it to a new $N+1^{th}$ does not yield accurate results. This is because the unified mapping function $\phi$ assigns equal importance to time-steps of different states, and (b) cannot be decomposed into various subspaces to collectively provide a new and unique representation of system states. These representations are crucial in potentially describing state dynamics for a $N+1^{th}$ System.

\begin{rmk}
Interestingly, the approach adopted in a TST model echoes the strategy employed in the OASIS framework by Kwon and colleagues \cite{bhadriraju2021oasis, narasingam2018data, Bhadriraju2019}. Essentially, the OASIS model can effectively characterize system dynamics, even with varying system parameters, such as changes in rate kinetics or mass transfer rate in a reactor system. This is similar to differing $G$ and $B$ values across $N$ various crystal systems. This feat is achieved by training several local SINDy models, followed by a supervisory DNN in the OASIS framework that learns the distinct parameters of the SINDy model for each $K^{th}$ local system. Although the OASIS framework is constructed by explicitly identifying $N$ different local SINDy models and subsequently combining them, the TST model emulates a similar behavior, albeit implicitly (without direct user intervention). This represents a stride toward the next generation of TL-based models. 
\end{rmk}

\begin{figure}[!ht]
	\begin{center} 
		\centerline{\includegraphics[width=0.9\columnwidth]{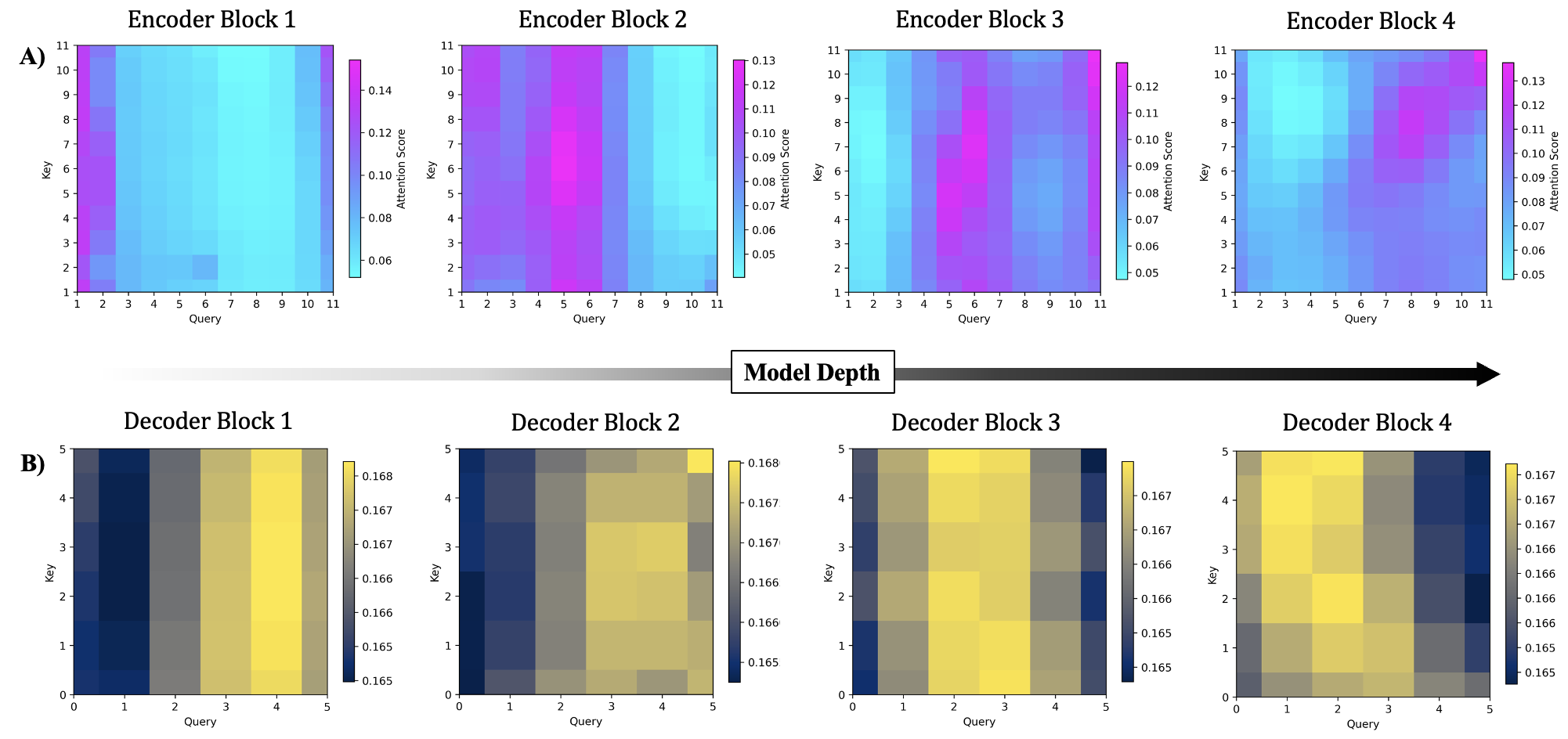}}
	\end{center}\vspace{10pt}
	\caption{A compilation of self-attention and cross-attention scores for encoders and decoders, respectively. These results correspond to the CrystalGPT-Base model with a window size of $W=12$. }\vspace{10pt}
	\label{attention_scores}
\end{figure}

\begin{figure}[!ht]
	\begin{center}
		\centerline{\includegraphics[width=0.8\columnwidth]{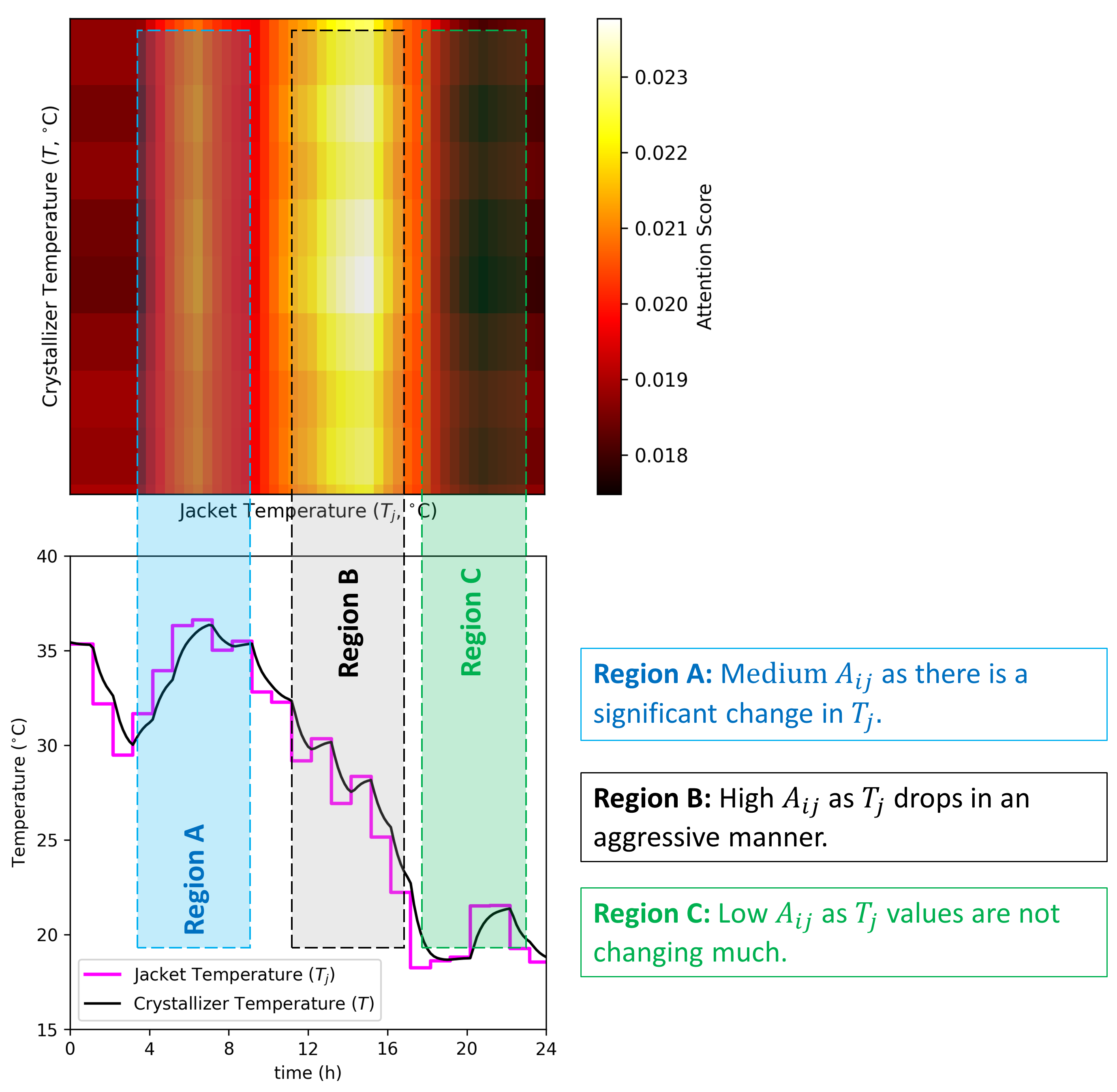}}
	\end{center}\vspace{10pt}
	\caption{The self-attentions scores between the jacket temperature input, $T_j$, and the crystallizer temperature, $T$. These results are associated with the CrystalGPT-Base model with a window size of $W=50$. $A_{ij}$ is the attention score.}\vspace{10pt}
	\label{temp_attn_scores}
\end{figure}

\subsection{Visualization of Attention Mechanism}

As previously discussed, the computation of attention scores within encoder and decoder blocks forms a crucial component of TSTs. In Figure~\ref{attention_scores}, we exhibit a compilation of attention scores across various encoders and decoders for the CrystalGPT-Base model configured with parameters $W\!=\!12$, $H\!=\!6$, and $\lambda\!=\!3$. In the encoder blocks, the attention scores illustrate the self-attention mechanism, empowering the TST to decipher the interplay among system dynamics within the source sequence. Notably, in shallower encoder blocks (e.g., Encoder Block 1), the attention scores appear evenly distributed but gradually skew towards the end of the sequence in deeper encoder blocks (e.g., in Encoder Block 4). This observation suggests that due to the nonlinear transformation occurring through the FFN in each encoder block $i$, a deeper encoder block $j$ exhibits more dispersed attention scores. These observations hint at the TST's capability to adeptly learn the nonlinear relationships governing the temporal evolution of system states. Similarly, the cross-attention scores in the decoder blocks (Figure~\ref{attention_scores}B) reveal intriguing patterns. In Decoder Block 1, the attention score is higher for regions that correspond to Queries with index [3,4,5]. This is due to the overlap ($\lambda = 3$) between the decoder input and the encoder input (Figure~\ref{TST_schematic}). Since the initial three Queries (i.e., corresponding to index [1,2,3]) on the decoder side ($X_{DEC}$) are identical to the final three Queries (i.e., corresponding to the index [3,4,5]) on the encoder side ($X_{ENC}$), the TST already knows them during its internal computation. On the contrary, TST is seeing the Queries with index [3,4,5] for the first time, and since its objective is to learn the continuity between encoder and decoder states, high attention score is awarded to this region. Lastly, similar to the case with encoders, successive deeper decoder blocks present a more intricate nonlinear relationship between \textbf{Q} and \textbf{V} values, as exemplified in Decoder Block 4. 

\begin{figure}[!ht]
	\begin{center}
		\centerline{\includegraphics[width=0.8\columnwidth]{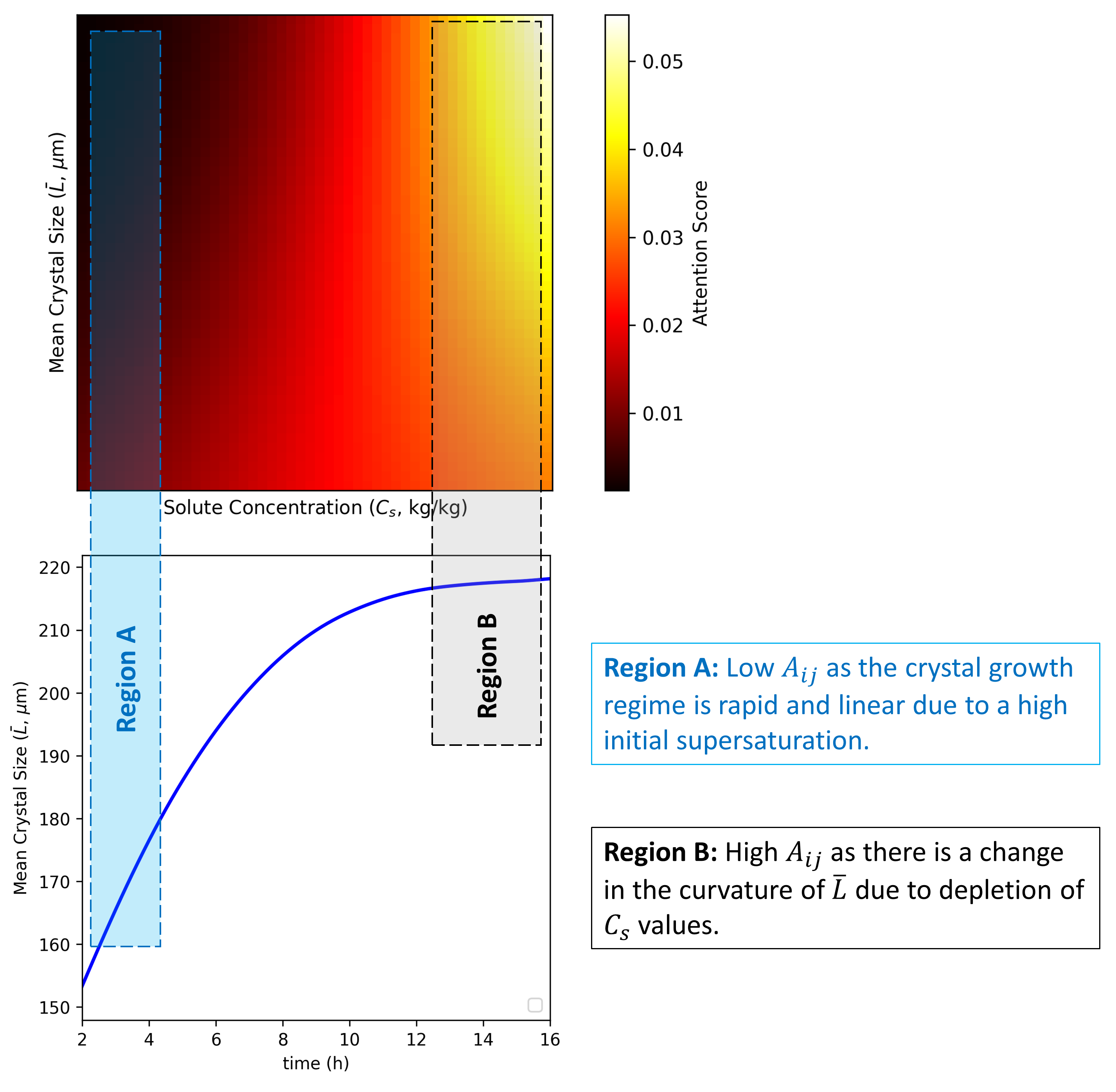}}
	\end{center}\vspace{10pt}
	\caption{Self-attentions scores between $C_s$ and $\bar{L}$. It is important to note that these are the results for CrystalGPT-Base with $W=50$, and $A_{ij}$ is the attention score.}\vspace{10pt}
	\label{conc_attn_scores}
\end{figure}

While Figure~\ref{attention_scores} can offer heuristic insights into the internal computations of TSTs, its attention scores are computed for the internal lifted states ($d_{model}$), which limits interpretability. Consequently, the same attention score mechanism has been reassessed within the original feature space, as depicted in Figures~\ref{temp_attn_scores} and \ref{conc_attn_scores}. Specifically, Figure~\ref{temp_attn_scores} shows variations in self-attention scores between the input jacket temperature ($T_j$) and the internal crystallizer temperature ($T$). Region A presents moderate attention scores, corresponding to step changes in the jacket temperature  profile ($T_j$). In contrast, Region B shows high attention scores when $T_j$ drops sharply, which directly impacts $T$ due to the energy balance equation outlined in Eq.~(\ref{MEBE}). Region C, where changes in $T_j$ are relatively subtle leading to minor shifts in $T$, results in lower attention scores. Likewise, in Figure~\ref{conc_attn_scores}, Region A represents the crystallizer's initial operation with a high initial concentration; the growth rate in this regime is relatively linear, thus yielding low attention scores. In contrast, Region B shows high attention scores due to a sudden curvature change in the evolution of $\bar{L}$, caused by the saturation of crystal size. Given the significance of this shift in the growth regime as a critical system characteristic. this region is assigned high attention scores.

\subsection{Real-world Applications of CrystalGPT: High-precision Model Predictions and MPC-based Setpoint Tracking}

Finally, to demonstrate CrystalGPT's practical relevance for various industrial tasks, we consider a mode prediction task and a setpoint tracking task utilizing MPC. 

\begin{figure}[!ht]
	\begin{center}
		\centerline{\includegraphics[width=1\columnwidth]{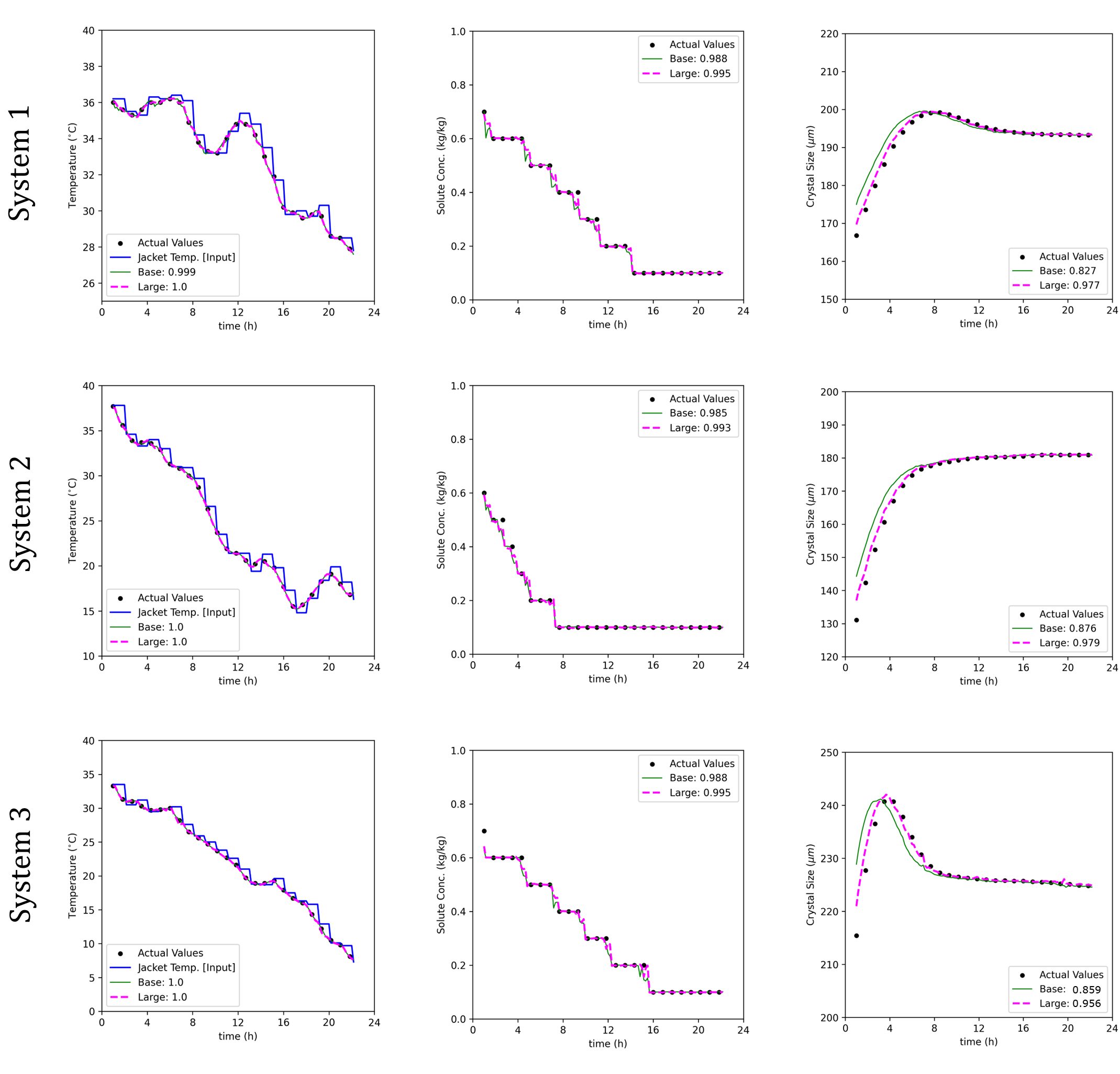}}
	\end{center}\vspace{10pt}
	\caption{Comparison of state predictions between CrystalGPT-Base and CrystalGPT-Large models for three arbitrary crystal systems.}\vspace{10pt}
	\label{CrystalGPT: Comparison Image}
\end{figure}

\subsubsection{Model Predictions with CrystalGPT}
We randomly selected three crystal systems from a set of 20 different systems and considered arbitrary operating conditions. The prediction results for these cases are illustrated in Figure~\ref{CrystalGPT: Comparison Image}. It is evident that the CrystalGPT-Large model yields an $R^2$ value exceeding 0.95 for all states across all cases, while the CrystalGPT-Base model achieves an $R^2$ value surpassing 0.85. It is noteworthy to mention that each of these crystal systems exhibits distinct dynamics. Despite CrystalGPT not being fine-tuned on any of these specific systems, it demonstrates exceptional predictive performance. For example, the crystal size evolution for System 2 shows rapid growth followed by saturation behavior. In contrast, System 3 shows a dampening effect in crystal size evolution, owing to an over-dominant nucleation rate that leads to the formation of crystal fines, thereby reducing the mean crystal size. With current SOTA ML models (e.g., LSTM and RNN), the generation of dedicated bespoke models would be necessary for Systems 2 and 3 due to their widely different kinetics. CrystalGPT, however, eliminates the need for such tailored and highly fine-tuned models while still maintaining high prediction accuracy. Furthermore, model inputs only include easily measurable process data (i.e., $T_j$, $T$, $C_s$, $\bar{L}$, $M_T$, and time) and do not incorporate any system-specific parameters (e.g., $G_i$ and $B_i$ or initial seeding amount and its CSD). Therefore, CrystalGPT relies solely on the state information of the current and past $W$ time-steps to make model predictions across any chosen crystal system. Remarkably, under these nontrivial circumstances, the model's predictions remain highly accurate across all three selected systems. 

\subsubsection{CrystalGPT-based MPC}

The preceding section highlights the impressive S2S transferability of CrystalGPT across a broad range of crystal systems and operating conditions, while maintaining highly precise time-series predictions. These predictive capabilities can be effectively utilized for online process control tasks, specifically for setpoint tracking across various crystal systems. To illustrate, we incorporated CrystalGPT within an MPC framework to perform setpoint tracking for the target crystal system, as shown below: 

\begin{equation} \label{optimizer}
	\begin{aligned}
		& \underset{T_{j}(t)}{Minimize}
		&&  \left (\bar{L} - L_{sp} \right)^2  \\
		&\text{s.t}
		&& T_{j,min} \leq T_j \leq T_{j,max} \\
		&&&  \delta T_{j,min} \leq \Delta T_j \leq \delta T_{j,max}  \\
		&&& \frac{\partial n}{\partial t} + \frac{\partial (Gn)}{\partial L} = 0 \\	
		&&&	\frac{d C_s}{d t} = -3\rho_ck_vG\mu_2\\
		&&& 	mC_p\frac{dT}{dt} = -UA(T - T_{j}) - \Delta H\rho_{}3k_vG\mu_2\\
		&&& G_i = a_{\scriptscriptstyle G} \exp\left(\frac{-b_{\scriptscriptstyle G}}{RT}\right) (S-1)^{\scriptscriptstyle C_{\scriptscriptstyle G}} \\
		&&&  B_i = a_B M_T^{b_B}(S-1)^{\scriptscriptstyle C_{\scriptscriptstyle B}}  \\ 
	\end{aligned}
\end{equation}
where $\bar{L}$ is the mean crystal size, $L_{sp}$ is the desired set-point, and $\Delta T_j$ corresponds to the maximum allowable change in jacket temperature between two sampling times. The MPC tracks the setpoint value of mean crystal size ($L_{sp}$) by manipulating $T_j$. In this setup, CrystalGPT serves as an internal surrogate model, with the PBM and MEBEs acting as a virtual experiment for state feedback. Furthermore, the MPC adopts a receding horizon strategy, where an entire trajectory of optimal inputs is initially calculated (e.g., a sequence of 10-step inputs). Here, only the first value of the optimal input is implemented, and as the process progresses toward the end time, fewer optimal inputs are computed until the final optimal input is determined. Overall, the MPC problem is formulated to minimize the deviation from the desired set-point mean crystal size while adhering to practical operating constraints.

\begin{figure}[!ht]
	\begin{center}
		\centerline{\includegraphics[width=1\columnwidth]{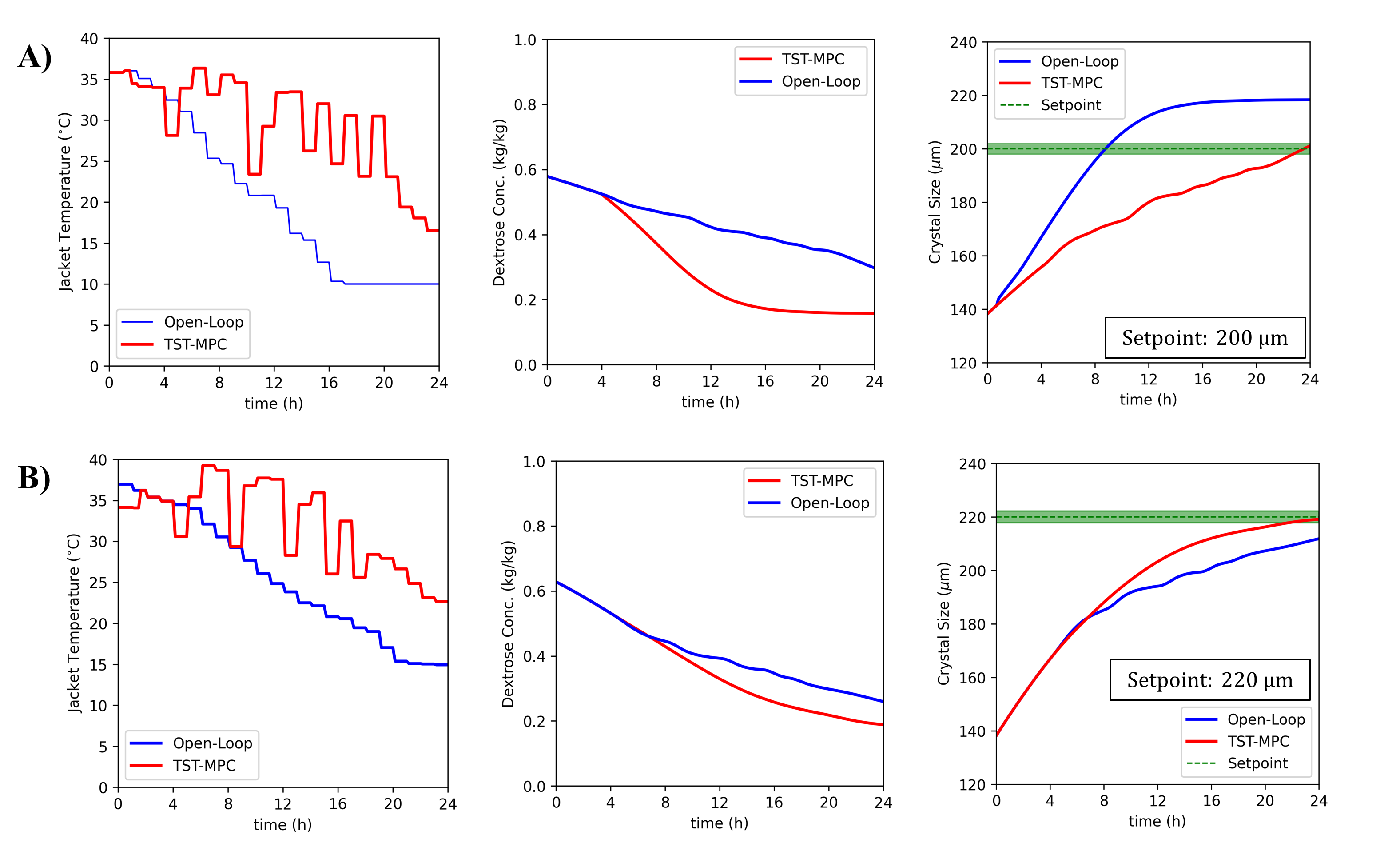}}
	\end{center}\vspace{5pt}
	\caption{Closed-loop results of the CrystalGPT-based MPC, applied to a variety of setpoint cases for a new crystal system, $N+1$. The green band denotes a setpoint deviation error margin of 1\%. }\vspace{10pt}
	\label{MPC_results}
\end{figure}

To further scrutinize the capabilities of CrystalGPT, setpoint tracking for the $21^{st}$ crystal system was simulated using the CrystalGPT-Large model, without any additional fine-tuning for this new system. For context, this $21^{st}$ System shares structural similarities in the $G$ and $B$ equations, PBM, and MEBEs with the previous systems, though its system parameters are chosen arbitrarily from the distribution displayed in Figure~\ref{variation_G_and_B}. Figure~\ref{MPC_results} illustrates the closed-loop simulation results of the CrystalGPT-based MPC for two different set-point tracking cases (i.e., 200$\mu$m and 220 $\mu$m). As can be seen from the figure, the MPC can effectively track the setpoint within a deviation range of less than 1\% (denoted by the green band in Figure~\ref{MPC_results}), whereas the open-loop result displays a deviation of 10\% to 15\%. Upon examination of the manipulated input profiles, it becomes apparent that the MPC adopts a strategic approach, manipulating the external jacket temperature ($T_j$) to regulate crystal growth. This is accomplished by generating alternating zones of high and low saturation. For instance, setting an intermediate low $T_j$ value results in decreased solute solubility and increased supersaturation, thereby facilitating rapid crystal growth. Conversely, employing a high intermediate $T_j$ value diminishes supersaturation and induces slower crystal growth, due to the proportional relationship it shares with growth kinetics. Furthermore, the solute concentration profile reveals an initial linear trend, followed by a sharp decline in concentration, indicative of the rapid crystal growth occurring in the early crystallization stages. Following this, the MPC adjusts the jacket temperature ($T_j$) as previously described, facilitating more controlled and steady crystal growth. This regulated growth pattern minimizes the depletion of solute concentration, leading to a gentler decline. Remarkably, all the above has been achieved without any additional training data from the $21^{st}$ System for the CrystalGPT-Large model, relying solely on its transfer learning capabilities to generate model predictions for ten future sampling times, thereby enabling the calculations for the MPC.

\section{Conclusions}
Existing digital twins exhibit limitations in their applicability across different chemical systems due to their system-specific design. This limitation presents a significant barrier to their broad-scale deployment. To overcome this issue, we devised a novel TST framework that demonstrates exceptional S2S transferability while maintaining precise time-series predictions. Specifically, we chose the batch crystallization of 20 unique crystal systems, derived from a set of sugar crystal systems, as a representative case study. We collected easily measurable process data (i.e., $T_j$, $C_s$, $\bar{L}$, etc.), compiling a massive dataset of approximately 10M+ datapoints. We used this dataset to train a TST model (CrystalGPT) that achieves an NMSE of $10^{-3}$, not only across the 20 distinct systems but also for an unseen $21^{st}$. In every case, the NMSE values for CrystalGPT are impressively 8 to 10 times lower than those for the SOTA LSTM model. Furthermore, we demonstrated the practical applicability of CrystalGPT for model predictions and online MPC implementation, achieving average $R^2$ values above 0.95 and a setpoint deviation of 1\%, respectively. In addition, we extensively explained the internal operations of the TST model. We hypothesize that the structural similarities across $N$ distinct systems (for example, $G$ and $B$ equations, PBM, and MEBEs) are captured by multiple attention heads. These attention heads serve as modular components (i.e., distributed models) that independently model various subspaces, each contributing to the comprehensive mapping between inputs and outputs. Moreover, the adaptability of the weighting coefficients for these attention heads enables seamless S2S transferability in TST models. Lastly, the TST framework we developed lays a robust foundation for constructing highly versatile digital twins that exhibit extraordinary S2S transferability for various modeling and control tasks. For example, similar models can be developed for different chemical systems, such as LionGPT for battery systems or TransFERMER for fermentation systems, among others. This opens up new avenues for extending the capabilities of digital twins across diverse applications.

\section{Acknowledgments}
Financial support from the Artie McFerrin Department of Chemical Engineering, and the Texas A\&M Energy Institute is gratefully acknowledged.

\newpage

%Bibliography
\bibliographystyle{unsrt}  
\bibliography{CrystalGPT}

\begin{thebibliography}{10}

\bibitem{bhadriraju2019machine}
Bhavana Bhadriraju, Abhinav Narasingam, and Joseph~S Kwon.
\newblock Machine learning-based adaptive model identification of systems:
  Application to a chemical process.
\newblock {\em Chemical Engineering Research and Design}, 152:372--383, 2019.

\bibitem{bhadriraju2021oasis}
Bhavana Bhadriraju, Joseph~Sang Kwon, and Faisal Khan.
\newblock {OASIS}-{P}: Operable adaptive sparse identification of systems for
  fault prognosis of chemical processes.
\newblock {\em Journal of Process Control}, 107:114--126, 2021.

\bibitem{sitapure2021multiscale}
Niranjan Sitapure, Robert Epps, Milad Abolhasani, and Joseph Sang-Il Kwon.
\newblock Multiscale modeling and optimal operation of millifluidic synthesis
  of perovskite quantum dots: towards size-controlled continuous manufacturing.
\newblock {\em Chemical Engineering Journal}, 413:127905, 2021.

\bibitem{sitapure2021cfd}
Niranjan Sitapure, Robert~W Epps, Milad Abolhasani, and Joseph~S Kwon.
\newblock {CFD}-based computational studies of quantum dot size control in slug
  flow crystallizers: Handling slug-to-slug variation.
\newblock {\em Industrial \& Engineering Chemistry Research},
  60(13):4930--4941, 2021.

\bibitem{zheng2022machine}
Yingzhe Zheng, Xiaonan Wang, and Zhe Wu.
\newblock Machine learning modeling and predictive control of the batch
  crystallization process.
\newblock {\em Industrial \& Engineering Chemistry Research},
  61(16):5578--5592, 2022.

\bibitem{lima2022development}
Fernando Arrais~RD Lima, Marcellus~GF de~Moraes, Argimiro~R Secchi, and
  Maur{\'\i}cio~B de~Souza~Jr.
\newblock Development of a recurrent neural networks-based {NMPC} for
  controlling the concentration of a crystallization process.
\newblock {\em Digital Chemical Engineering}, 5:100052, 2022.

\bibitem{sitapure2022neural}
Niranjan Sitapure and Joseph Sang-Il Kwon.
\newblock Neural network-based model predictive control for thin-film chemical
  deposition of quantum dots using data from a multiscale simulation.
\newblock {\em Chemical Engineering Research and Design}, 183:595--607, 2022.

\bibitem{hwang2022model}
Gyuyeong Hwang, Niranjan Sitapure, Jiyoung Moon, Hyeonggeon Lee, Sungwon Hwang,
  and Joseph~Sang Kwon.
\newblock Model predictive control of lithium-ion batteries: Development of
  optimal charging profile for reduced intracycle capacity fade using an
  enhanced single particle model ({SPM}) with first-principled
  chemical/mechanical degradation mechanisms.
\newblock {\em Chemical Engineering Journal}, 435:134768, 2022.

\bibitem{bangi2020deep}
Mohammed Saad~Faizan Bangi and Joseph~S Kwon.
\newblock Deep hybrid modeling of chemical process: application to hydraulic
  fracturing.
\newblock {\em Computers \& Chemical Engineering}, 134:106696, 2020.

\bibitem{bangi2022physics}
Mohammed Saad~Faizan Bangi, Katy Kao, and Joseph~Sang Kwon.
\newblock Physics-informed neural networks for hybrid modeling of lab-scale
  batch fermentation for $\beta$-carotene production using saccharomyces
  cerevisiae.
\newblock {\em Chemical Engineering Research and Design}, 179:415--423, 2022.

\bibitem{shah2022deep}
Parth Shah, M~Ziyan Sheriff, Mohammed Saad~Faizan Bangi, Costas Kravaris,
  Joseph~Sang Kwon, Chiranjivi Botre, and Junichi Hirota.
\newblock Deep neural network-based hybrid modeling and experimental validation
  for an industry-scale fermentation process: Identification of time-varying
  dependencies among parameters.
\newblock {\em Chemical Engineering Journal}, 441:135643, 2022.

\bibitem{vaswani2017attention}
Ashish Vaswani, Noam Shazeer, Niki Parmar, Jakob Uszkoreit, Llion Jones,
  Aidan~N Gomez, {\L}ukasz Kaiser, and Illia Polosukhin.
\newblock Attention is all you need.
\newblock {\em Advances in Neural Information Processing Systems}, 30, 2017.

\bibitem{devlin2018bert}
Jacob Devlin, Ming-Wei Chang, Kenton Lee, and Kristina Toutanova.
\newblock {BERT}: Pre-training of deep bidirectional transformers for language
  understanding.
\newblock {\em arXiv preprint}, 1810.04805, 2018.

\bibitem{shoeybi2019megatron}
Mohammad Shoeybi, Mostofa Patwary, Raul Puri, Patrick LeGresley, Jared Casper,
  and Bryan Catanzaro.
\newblock Megatron-{LM}: Training multi-billion parameter language models using
  model parallelism.
\newblock {\em arXiv preprint}, 1909.08053, 2019.

\bibitem{radford2019language}
Alec Radford, Jeffrey Wu, Rewon Child, David Luan, Dario Amodei, and Ilya
  Sutskever.
\newblock Language models are unsupervised multitask learners.
\newblock {\em OpenAI blog}, 1(8):9, 2019.

\bibitem{brown2020language}
Tom Brown, Benjamin Mann, Nick Ryder, Melanie Subbiah, Jared~D Kaplan, Prafulla
  Dhariwal, Arvind Neelakantan, Pranav Shyam, Girish Sastry, Amanda Askell,
  et~al.
\newblock Language models are few-shot learners.
\newblock {\em Advances in Neural Information Processing Systems},
  33:1877--1901, 2020.

\bibitem{liu2021swin}
Ze~Liu, Yutong Lin, Yue Cao, Han Hu, Yixuan Wei, Zheng Zhang, Stephen Lin, and
  Baining Guo.
\newblock Swin transformer: Hierarchical vision transformer using shifted
  windows.
\newblock In {\em Proceedings of the IEEE/CVF International Conference on
  Computer Vision}, pages 10012--10022, 2021.

\bibitem{rombach2022high}
Robin Rombach, Andreas Blattmann, Dominik Lorenz, Patrick Esser, and Bj{\"o}rn
  Ommer.
\newblock High-resolution image synthesis with latent diffusion models.
\newblock In {\em Proceedings of the IEEE/CVF Conference on Computer Vision and
  Pattern Recognition}, pages 10684--10695, 2022.

\bibitem{vogel2023learning}
Gabriel Vogel, Lukas~Schulze Balhorn, and Artur~M Schweidtmann.
\newblock Learning from flowsheets: A generative transformer model for
  autocompletion of flowsheets.
\newblock {\em Computers \& Chemical Engineering}, page 108162, 2023.

\bibitem{kang2023multi}
Yeonghun Kang, Hyunsoo Park, Berend Smit, and Jihan Kim.
\newblock A multi-modal pre-training transformer for universal transfer
  learning in metal-organic frameworks.
\newblock {\em Nature Machine Intelligence}, 5:2522--5839, 2023.

\bibitem{mann2021predicting}
Vipul Mann and Venkat Venkatasubramanian.
\newblock Predicting chemical reaction outcomes: A grammar ontology-based
  transformer framework.
\newblock {\em AIChE Journal}, 67(3):e17190, 2021.

\bibitem{Clauwaert2021Explainability}
Jim Clauwaert, Gerben Menschaert, and Willem Waegeman.
\newblock Explainability in transformer models for functional genomics.
\newblock {\em Briefings in Bioinformatics}, 22(5):060, 2021.

\bibitem{yun2019transformers}
Chulhee Yun, Srinadh Bhojanapalli, Ankit~Singh Rawat, Sashank~J Reddi, and
  Sanjiv Kumar.
\newblock Are transformers universal approximators of sequence-to-sequence
  functions?
\newblock {\em arXiv preprint}, 1912:10077, 2019.

\bibitem{wen2022transformers}
Qingsong Wen, Tian Zhou, Chaoli Zhang, Weiqi Chen, Ziqing Ma, Junchi Yan, and
  Liang Sun.
\newblock Transformers in time series: A survey.
\newblock {\em arXiv preprint}, 2202.07125, 2022.

\bibitem{sitapure2023exploring}
Niranjan Sitapure and Joseph Sang-Il Kwon.
\newblock Exploring the potential of time-series transformers for process
  modeling and control in chemical systems: an inevitable paradigm shift?
\newblock {\em Chemical Engineering Research and Design}, 194:461--477, 2023.

\bibitem{zeng2022transformers}
Ailing Zeng, Muxi Chen, Lei Zhang, and Qiang Xu.
\newblock Are transformers effective for time series forecasting?
\newblock {\em arXiv preprint}, 2205:13504, 2022.

\bibitem{shi1990crystallization}
Y~Shi, B~Liang, and RW~Hartel.
\newblock Crystallization kinetics of alpha-lactose monohydrate in a continuous
  cooling crystallizer.
\newblock {\em Journal of Food science}, 55(3):817--820, 1990.

\bibitem{ouiazzane2008estimation}
S~Ouiazzane, B~Messnaoui, S~Abderafi, J~Wouters, and T~Bounahmidi.
\newblock Estimation of sucrose crystallization kinetics from batch
  crystallizer data.
\newblock {\em Journal of Crystal Growth}, 310(4):798--803, 2008.

\bibitem{markande2012influence}
Abhay Markande, Amale Nezzal, John Fitzpatrick, Luc Aerts, and Andreas Redl.
\newblock Influence of impurities on the crystallization of dextrose
  monohydrate.
\newblock {\em Journal of Crystal Growth}, 353(1):145--151, 2012.

\bibitem{sitapure2023unified}
Niranjan Sitapure and Joseph~Sang Kwon.
\newblock A unified approach for modeling and control of crystallization of
  quantum dots ({QD}s).
\newblock {\em Digital Chemical Engineering}, 6:100077, 2023.

\bibitem{worlitschek2004model}
J{\"o}rg Worlitschek and Marco Mazzotti.
\newblock Model-based optimization of particle size distribution in
  batch-cooling crystallization of paracetamol.
\newblock {\em Crystal Growth \& Design}, 4(5):891--903, 2004.

\bibitem{epps2017automated}
Robert~W Epps, Kobi~C Felton, Connor~W Coley, and Milad Abolhasani.
\newblock Automated microfluidic platform for systematic studies of colloidal
  perovskite nanocrystals: towards continuous nano-manufacturing.
\newblock {\em Lab on a Chip}, 17(23):4040--4047, 2017.

\bibitem{dong2018precise}
Yitong Dong, Tian Qiao, Doyun Kim, David Parobek, Daniel Rossi, and Dong~Hee
  Son.
\newblock Precise control of quantum confinement in cesium lead halide
  perovskite quantum dots via thermodynamic equilibrium.
\newblock {\em Nano Letters}, 18(6):3716--3722, 2018.

\bibitem{sitapure2020kinetic}
Niranjan Sitapure, Tian Qiao, Dong Son, and Joseph~S Kwon.
\newblock Kinetic monte carlo modeling of the equilibrium-based size control of
  {C}s{P}b{B}r$_3$ perovskite quantum dots in strongly confined regime.
\newblock {\em Computers \& Chemical Engineering}, 139:106872, 2020.

\bibitem{Levine2020Limits}
Yoav Levine, Noam Wies, Or~Sharir, Hofit Bata, and Amnon Shashua.
\newblock Limits to depth efficiencies of self-attention.
\newblock {\em Advances in Neural Information Processing Systems},
  33:22640--22651, 2020.

\bibitem{Xu2019Lipschitz}
Hongfei Xu, Qiuhui Liu, Josef van Genabith, Deyi Xiong, and Jingyi Zhang.
\newblock Lipschitz constrained parameter initialization for deep transformers.
\newblock {\em arXiv preprint}, 1911:03179, 2019.

\bibitem{song2020specswap}
Xingcheng Song, Zhiyong Wu, Yiheng Huang, Dan Su, and Helen Meng.
\newblock Specswap: A simple data augmentation method for end-to-end speech
  recognition.
\newblock In {\em Interspeech 2020, Shanghai, China}, pages 581--585, 2020.

\bibitem{narasingam2018data}
Abhinav Narasingam and Joseph~S Kwon.
\newblock Data-driven identification of interpretable reduced-order models
  using sparse regression.
\newblock {\em Computers \& Chemical Engineering}, 119:101--111, 2018.

\bibitem{Bhadriraju2019}
Bhavana Bhadriraju, Abhinav Narasingam, and Joseph~Sang Kwon.
\newblock Machine learning-based adaptive model identification of systems:
  application to a chemical process.
\newblock {\em Chemical Engineering Research and Design}, 152:372--383, 2019.

\end{thebibliography}

\end{document}